\documentclass[11pt,a4paper]{article}
\pdfoutput=1
\usepackage[utf8]{inputenc}
\usepackage{jheppub}
\allowdisplaybreaks

\def\be{\begin{equation}}
\def\ee{\end{equation}}
\def\bea{\begin{eqnarray}}
\def\eea{\end{eqnarray}}
\def\beal{\begin{equation}\begin{aligned}}
\def\eeal{\end{aligned}\end{equation}}
\def\bse{\begin{subequations}}
\def\ese{\end{subequations}}
\def\nn{\nonumber}

\def\fla{r}
\def\flb{{\hat r}}

\def\bra#1{\langle #1|}
\def\ket#1{|#1 \rangle}

\def\braket#1{\langle #1 \rangle}

\def\u#1{\underline{#1}}
\def\o#1{\overline{#1}}

\def\la{\lambda}
\def\lb{\tilde{\lambda}}

\def\Res_#1{\operatorname*{Res}_{#1}}
\def\sgn{\operatorname*{sgn}}
\def\Tr{\operatorname*{Tr}}

\def\d{\mathrm{d}}
\def\dd{d\!\!{}^-\!}
\def\del{\delta\!\!\!{}^-\!}

\def\cN{\mathcal{N}}
\def\gg{\mathfrak{g}}
\def\hh{\mathfrak{h}}
\def\tf{\tilde{f}}

\def\ie{i.e. }
\def\eg{e.g. }

\def\eqn#1{eq.~\eqref{#1}}
\def\eqns#1#2{eqs.~\eqref{#1} and~\eqref{#2}}

\def\Eqn#1{Eq.~\eqref{#1}}

\def\tab#1{Table~{\ref{#1}}}

\def\sec#1{section~{\ref{#1}}}

\def\secss#1#2#3{sections~{\ref{#1}}, {\ref{#2}} and~{\ref{#3}}}

\def\app#1{appendix~{\ref{#1}}}

\def\rcite#1{ref.~\cite{#1}}
\def\rcites#1{refs.~\cite{#1}}

\def\Rcites#1{Refs.~\cite{#1}}

\newcommand{\scalegraph}[2]{\vcenter{\hbox{\!\;\includegraphics[scale=#1]{graphs/#2.pdf}\!\;}}}

\title{Double copy for massive quantum particles with spin}

\author[a,b]{Henrik Johansson}
\author[c]{and Alexander Ochirov}

\affiliation[a]{Department of Physics and Astronomy, Uppsala University, SE-75108 Uppsala, Sweden}
\affiliation[b]{Nordita, Stockholm University and KTH Royal Institute of Technology and Stockholm University, Roslagstullsbacken 23, SE-10691 Stockholm, Sweden}
\affiliation[c]{ETH Z\"urich, Institut f\"ur Theoretische Physik,
Wolfgang-Pauli-Str. 27, 8093 Z\"urich, Switzerland}

\emailAdd{henrik.johansson@physics.uu.se,aochirov@phys.ethz.ch}

\preprint{UUITP-24/19 \\
\phantom{~} \hfill NORDITA 2019-070}

\abstract{The duality between color and kinematics was originally observed for purely adjoint massless gauge theories, and later found to hold even after introducing massive fermionic and scalar matter in arbitrary gauge-group representations. Such a generalization was critical for obtaining both loop amplitudes in pure Einstein gravity and realistic gravitational matter from the double copy. In this paper we elaborate on the double copy that yields amplitudes in gravitational theories coupled to flavored massive matter with spin, which is relevant to the problems of black-hole scattering and gravitational waves. Our construction benefits from making the little group explicit for the massive particles, as shown on lower-point examples. For concreteness, we focus on the double copy of QCD with massive quarks, for which we work out the gravitational Lagrangian up to quartic scalar and vector-scalar couplings. We find new gauge-invariant double-copy formulae for tree-level amplitudes with two distinct-flavor pairs of matter and any number of gravitons. These are similar to, but inherently different from, the well-known Kawai-Lewellen-Tye formulae, since the latter only hold for the double copy of purely adjoint gauge theories.
}

\begin{document}

\maketitle

\section{Introduction}
\label{sec:intro}

With the discovery of gravitational waves
by the LIGO and VIRGO experiments~\cite{Abbott:2016blz,TheLIGOScientific:2017qsa},
we are now entering an era where observations and theory
will join forces to explore gravitational physics at unprecedented scales.
Future experiments are in the planning, and on the theory side
calculation methods are being sharpened to handle the need for higher precision.
Focusing on analytical methods that use scattering amplitudes,
there are by now a number of impressive results and techniques
relevant to black-hole scattering~\cite{Neill:2013wsa,Vaidya:2014kza,
Bjerrum-Bohr:2018xdl,Cheung:2018wkq,Kosower:2018adc,Guevara:2018wpp,Bern:2019nnu,
Brandhuber:2019qpg,Cristofoli:2019neg,Maybee:2019jus,Guevara:2019fsj}.
A common theme of this work is the observation that
gravitational amplitudes are vastly simpler than any Lagrangian may suggest.
Together with the general expectation that all physical information
of a theory should be contained in its S-matrix,
one may hope to circumvent traditional methods
and extract the needed quantities directly from amplitudes.

One of the most remarkable properties of gravitational scattering amplitudes
is a double-copy structure that allows them to be factorized in terms of simpler gauge-theory amplitudes. It was first observed by Kawai, Lewellen and Tye (KLT)~\cite{Kawai:1985xq}
in the context of closed-string amplitudes, and it became a systematic method
applicable at loop level and for a multitude of theories with the introduction of the Bern-Carrasco-Johansson (BCJ) double copy~\cite{Bern:2008qj,Bern:2010ue}.
The BCJ double copy identified the underlying gauge-theory property --- a duality
between color and kinematics~\cite{Bern:2008qj,Bern:2010ue}.
This duality is present in many physically relevant gauge theories,
such as pure Yang-Mills theory~\cite{Bern:2008qj,BjerrumBohr:2009rd, Stieberger:2009hq,Bern:2013yya},
quantum chromodynamics
(QCD)~\cite{Johansson:2014zca,Johansson:2015oia,delaCruz:2015dpa},
Yang-Mills-Higgs models~\cite{Chiodaroli:2015rdg},
and supersymmetric generalizations thereof~\cite{Bern:2010ue,Bern:2011rj,
BoucherVeronneau:2011qv,Bern:2012uf,Bern:2012cd,Johansson:2014zca,
Chiodaroli:2015rdg,Johansson:2017bfl}.
The color-kinematics duality provides a mechanism that extends
gauge invariance of two gauge theories to a diffeomorphism symmetry
of the resulting gravity~\cite{Bern:2008qj,Monteiro:2011pc,Chiodaroli:2017ngp,
Chiodaroli:2016jqw,Arkani-Hamed:2016rak,Anastasiou:2018rdx}.
The  mechanism lends support to the expectation that the double copy may be a general feature of gravitational theories~\cite{Chiodaroli:2015wal,Chiodaroli:2017ehv}.
Indeed, by now it has been observed to give correct amplitudes
in a large set of gravitational theories,
such as pure Einstein gravity~\cite{Johansson:2014zca},
pure and matter-coupled supergravities~\cite{Bern:2010ue,Carrasco:2012ca,
Johansson:2014zca,Johansson:2017bfl,Ben-Shahar:2018uie},
Einstein-Yang-Mills theories~\cite{Chiodaroli:2014xia,Chiodaroli:2015rdg,
Chiodaroli:2015wal,Chiodaroli:2017ehv}, and many others.

Apart from flat-space scattering amplitudes, the double-copy structure has been found in other gravitational settings, such as for non-trivial curved backgrounds~\cite{Adamo:2018mpq,Farrow:2018yni}. Remarkably, a number of important classical solutions
to Einstein's field equations --- including the Schwarzschild and Kerr black holes --- may be expressed in terms of solutions
in the underlying gauge theory~\cite{Monteiro:2014cda,Luna:2015paa,Luna:2016hge,
Luna:2018dpt,Berman:2018hwd,Lee:2018gxc,Gurses:2018ckx,Cho:2019ype,Arkani-Hamed:2019ymq}.
Such a classical double copy,
introduced by Monteiro, O’Connell and White~\cite{Monteiro:2014cda},
relies on the metric parametrizations
in which Einstein's equations effectively linearize.
For genuinely non-linear problems, it
has been perturbatively extended to problems
involving classical sources and gravitational radiation~\cite{Luna:2016due,Goldberger:2016iau,Luna:2016hge,Bahjat-Abbas:2017htu,Carrillo-Gonzalez:2017iyj,Luna:2017dtq,Goldberger:2017vcg,
Goldberger:2017ogt,Shen:2018ebu,Plefka:2018dpa,Carrillo-Gonzalez:2018pjk,
Bahjat-Abbas:2018vgo,CarrilloGonzalez:2019gof,Bautista:2019tdr}.
This gives hope that the double copy
can become a method for state-of-the-art calculations
related to gravitational waves and effective black-hole potentials.
In fact, the amplitude-level double copy 
has already played a crucial role in the 
calculation of the
effective potential between two non-spinning black holes
to third order in the post-Minkowskian expansion~\cite{Bern:2019nnu}.

The classical double copy is still less well understood
than the original ``quantum double copy'' used for scattering amplitudes.
The precise prescriptive details are still under development,
and a number of puzzles remain.\footnote{Curiously,
a recent classical double-copy calculation of the effective action
of two massive spinless particles failed to match
the known next-to-next-to-leading order result~\cite{Plefka:2019hmz}.
It remains to be seen if this is because of some inherent problem,
or simply due to the ambiguities present in off-shell unphysical quantities.}
For instance, there is the issue of eliminating
the dilaton and axion from the double copy with sources,
which has not been solved in an entirely satisfactory manner.
In quantum theories without sources,
these two massless particles can be systematically canceled from loop diagrams
using matter that obey ghost statistics~\cite{Johansson:2014zca,Johansson:2017bfl}. 
In the presence of massive sources, however,
the mass parameter induces a linear coupling to the dilaton
(for scalar sources) and axion (for sources with spin).
Then it becomes non-trivial to consistently truncate
these massless particles even at the classical level.
Nevertheless, if one has access to the on-shell states, as in a unitarity cut,
it should be possible to always consistently project out
the unwanted states from the double copy~\cite{Bern:2019nnu}.
Alternatively, one can resort to computing the much simpler Feynman diagrams
that involve the undesired scalars,
and subtract them from the double copy~\cite{Luna:2017dtq}.
The latter approach resembles the ghost prescription at loop level,
but is still unclear if it is part of a systematic and simple double-copy formalism.

These puzzles call for more systematic studies
where classical perturbative results can be matched
to the corresponding amplitude calculations,
for which the double copy is better understood.
An interesting study in this spirit is \rcite{Shen:2018ebu} by Shen,
where a five-point amplitude,
directly related to the emission of gravitational radiation from binary system,
was computed in the classical limit at the next-to-leading post-Minkowskian order.
Notably, the calculation was simplified by mapping
the one-loop two-source problem to a tree-level amplitude with three sources.
It can thus be fruitful to improve our understanding of
tree-level gravity with more than two sources as obtained through the double copy.

The classical double copy in the presence of multiple sources
is indirectly related to earlier studies of a double copy for (super)gravity amplitudes
coupled to distinctly flavored matter particles.
In \rcite{Johansson:2015oia} the double-copy formula
\be
   {\cal M}_{n,k}^{\text{tree}} = \sum_{\sigma\in \text{basis}}
      K(1,2,\sigma) A(1,2,\sigma)\, ,
\label{DCformula2015}
\ee
was introduced for general matter-coupled tree amplitudes in gravity.
It takes as input gauge-theory partial amplitudes $A(1,2,\sigma)$ with distinctly flavored matter, such as QCD amplitudes, as well as kinematic factors $K(1,2,\sigma)$ that are dual to a basis of QCD color factors. This general construction bears strong resemblance to Shen's detailed calculation~\cite{Shen:2018ebu}. While it is applicable to tree-level gravitational processes for any number of sources, the formula has the drawback that the individual kinematic factors are gauge-dependent.
The standard KLT formulae~\cite{Kawai:1985xq,Bern:1998sv, BjerrumBohr:2010hn}, which are manifestly gauge-invariant, cannot be used together with QCD amplitudes that contain multiple quark pairs, since the fundamental quarks alter the general amplitude properties, such as the BCJ relations~\cite{Johansson:2015oia}.
Indeed, the KLT double copy implicitly assumes that the gauge theory has only adjoint particles, and if not, then the gauge-invariant double copy needs to take a different form.

In \rcite{delaCruz:2016wbr}
a first attempt was made to find a gauge-invariant double copy,
where a general formula applicable for QCD amplitudes was given in terms of an unknown KLT-like matrix.
Similarly to the adjoint case, it is related
to the inverse~\cite{Cachazo:2013iea} of the matrix
of the doubly color-stripped amplitudes
in a scalar theory with $\phi^3$ interactions
\cite{Du:2011js,BjerrumBohr:2012mg,Arkani-Hamed:2017mur,Bahjat-Abbas:2018vgo}.
The difference is that in the flavored case there are several types of scalars:
one bi-adjoint and a family of bi-fundamental scalars~\cite{Brown:2018wss}.
Since the tree amplitudes of the cubic scalar theory are in principle straightforward to work out
(see \eg \rcite{Mafra:2016ltu} and \sec{sec:kltberendsgiele} here),
the non-trivial problem is to invert the resulting matrix.
Its rank is determined by the number of partial QCD amplitudes
that are independent under the BCJ amplitude relations,
which at multiplicity $n$, for $k \ge 2$ sources, is given by $(n-3)!(2k-2)/k!$~\cite{Johansson:2015oia}.
In \sec{sec:bcj2klt}, we will determine this KLT-like matrix
for certain sectors in the $(n,k)$ space.
In particular, we give all-multiplicity formulae for $k \le 2$,
as well as lower-multiplicity results for $k=3$.

\paragraph{Identifying the gravitational theory.}

In this paper, we set out to provide a general framework
for massive double copies with spin.
We use the double copy of QCD as our prime example.
Picking a gauge-group representation of matter particles other than the adjoint
(\eg the fundamental representation of ${\rm SU}(N_c)$ for quarks)
makes it possible to separate matter and gluons in the double copy~\cite{Chiodaroli:2013upa,Johansson:2014zca}.
While the construction is guaranteed to give diffeomorphism-invariant amplitudes~\cite{Chiodaroli:2017ngp}, it is in general a non-trivial problem to specify the gravitational theory that results from the double copy.
The simplest step is, of course, to identify the spectrum of on-shell states.
For massless quarks, the QCD double copy produces either massless scalars or massless vectors
depending on how the spins of the quarks are aligned~\cite{Johansson:2014zca}.
However, for more realistic matter we take the quarks to be massive,
and then the double copy produces all the states
in the tensor product of two spin-1/2 particles:
a massive scalar and a massive abelian vector.
If we consider a family of $N_f$ quarks,
we obtain a family of flavored massive scalars and vectors
in the gravitational theory.
Note that each vector is by construction paired up with a scalar,
and they have the same mass as the quark from which they originate.
In addition to the massive matter,
we also have a massless dilaton and an axion present
in the gravitational theory, as reviewed in \app{app:fatgravity}.
See \tab{tab:particlecontent}
for a summary of the different constructions and state countings.

\begin{table*}
\centering
\begin{tabular}{|l|l|}
\hline 
Gauge theory &
Gravitational theory = (Gauge theory)$^2$ \\
\hline
YM &
Axiodilaton gravity~=~GR + dilaton + axion \\
$A^\mu$ &
$\hh^{\mu\nu} \oplus Z \oplus \bar{Z}$ \\
\hline
YM~+~$N_f$ Majorana &
Axiodilaton gravity~+~$N_f$ \{real scalars \& vectors\} \\
$A^\mu~\,\oplus\;\!N_f \times \Psi_{\rm M}$ &
$ \hh^{\mu\nu} \oplus Z \oplus \bar{Z}\qquad\quad\:\oplus\;\!N_f
  \times\{\varphi \oplus V^\mu\} $ \\
\hline
YM~+~$N_f$ Dirac~=~QCD &
Axiodilaton gravity~+~$N_f$ \{complex scalars \& vectors\} \\
$A^\mu~\,\oplus\;\!N_f \times\{\Psi \oplus \bar{\Psi}\}$ &
$ \hh^{\mu\nu} \oplus Z \oplus \bar{Z}\qquad\quad\:\oplus\;\!N_f
  \times\{\varphi \oplus \varphi^* \oplus V^\mu \oplus V^{*\mu}\} $\!\\
\hline
\end{tabular}
\caption[a]{\small Field content for pure Yang-Mills theory,
self-conjugate QCD~\cite{Anber:2019nfu} and conventional QCD and their respective double copies.}
\label{tab:particlecontent}
\end{table*}

A much harder problem is to understand how the gravitational matter interacts
in the double-copy theory.
For the massless and supersymmetric cases, one can identify
the resulting gravitational theories with certain minimally-matter-coupled
Maxwell-Einstein supergravities~\cite{Chiodaroli:2014xia,Chiodaroli:2015wal,Chiodaroli:2016jqw},
such as the ``generic Jordan family'' (in 6D)~\cite{Gunaydin:1983bi},
the ``generic non-Jordan family'' (in 5D)~\cite{Gunaydin:1986fg}
and the Luciani model (in 4D)~\cite{Luciani:1977hp}.
More precisely, the double copy  $(\text{QCD})\otimes(\cN=2~\text{SQCD})$
between two four-dimensional massless theories gives amplitudes
in the $\cN=2$ Luciani model~\cite{Chiodaroli:2015wal,Ben-Shahar:2018uie}. Truncating away the $\cN=2$ supersymmetry
should give the double copy $(\text{QCD})\otimes(\text{QCD})$
albeit in the massless case.

In \secss{sec:wavefunctions}{sec:3pt}{sec:4pt},
we work out much of the structure of the Lagrangian that
results from double copy $(\text{QCD})\otimes(\text{QCD})$,
by matching a Lagrangian ansatz with the amplitudes from the double copy.
To construct the amplitudes,
we exploit the little-group structure of the massive particles
by using the massive spinor-helicity formalism~\cite{Arkani-Hamed:2017jhn}.
The resulting Lagrangian inherits non-trivial matter interactions
from the supersymmetric parent theory.
However, the non-zero masses make the interactions
considerably more complicated than in the Luciani model~\cite{Luciani:1977hp}.
See \rcites{Chiodaroli:2017ehv,Chiodaroli:2018dbu} for other recent massive double-copy construction for gauged supergravity theories which may offer some guidance to the massive  Lagrangian. 

The gravitational Lagrangian has to satisfy certain properties
in order to be compatible with the double copy $(\text{QCD})\otimes(\text{QCD})$.
First, for $N_f=0$
the Lagrangian should match Einstein gravity coupled to a dilaton and axion,
called ``axiodilaton gravity'' henceforth.
The 4D renormalizability of QCD imposes, by dimensional analysis,
that the resulting gravitational Lagrangian have at most two derivatives in each term. Similarly, masses can appear at most quadratically,
and each power of~$m$ lowers the number of derivatives by one.
Moreover, the massive vector fields contain longitudinal modes,
which in the massless limit behave as $\varepsilon_\text{L}^\mu \sim p^\mu/m$.
Since they must produce well-behaved scalar interactions in the massless limit,
each explicit vector field $V_\mu$
must be multiplied by one power of $m$, or, alternatively,
be part of a covariant field strength $V_{\mu\nu} = 2\partial_{[\mu} V_{\nu]}$
that is insensitive to $V^\mu \sim p^\mu$.
By dimensional analysis,
this limits the vector fields to appear at most quadratically in the Lagrangian.
Finally, one should expect to obtain interactions consistent with a truncation of the Luciani model~\cite{Luciani:1977hp} in the massless limit.
However, this is quite a delicate matching, both because of the needed supersymmetric truncation and
since the longitudinal vector mode should non-trivially combine
with the scalar modes to form the complex scalars of the Luciani model.
More work is required to work out the details of this last matching. 

For convenience of the reader, below we quote all the determined terms
in the gravitational Lagrangian
obtained from the double copy of QCD,

{\fontsize{10.5pt}{0}\selectfont
\begin{align}
   {\cal L}_{\text{(QCD)}^2} =\,&
    - \frac{2}{\kappa^2} R
    + \frac{ \partial_\mu \bar{Z}\;\!\partial^\mu Z }
           { \big(1-\frac{\kappa^2\!}{4}\bar{Z}Z\big)^2 }
    + \sum_{\fla=1}^{N_f} \bigg\{
   {-}\frac{1}{2} V_{\fla \mu\nu}^* V_\fla^{\mu\nu}\!
    + m_\fla^2 V_{\fla \mu}^* V_\fla^\mu\:\!\!
      \bigg[ 1\:\!\!-\:\!\!\frac{\kappa}{2}(Z\:\!\!+\:\!\!\bar{Z})\:\!\!
           +\:\!\!\frac{\kappa^2\!}{2} \bar{Z} Z \bigg] \nn \\* &
    + \partial_\mu \varphi_\fla^*\;\!\partial^\mu \varphi_\fla\:\!\!
    - m_\fla^2 \varphi_\fla^* \varphi_\fla
      \bigg[ 1 - \frac{\kappa}{2}(Z\:\!\!+\:\!\!\bar{Z})
           + \frac{\kappa^2\!}{16} (Z^2\!+\:\!\!\bar{Z}^2\!+8\bar{Z} Z)
      \bigg] \nn \\* &
    + \frac{i\kappa}{4} m_\fla
      \Big[
      (\varphi_\fla^* V_{\fla \mu}\!+ V^*_{\fla \mu} \varphi_\fla)
      \partial^\mu (Z\:\!\!-\:\!\!\bar{Z})
    - (Z\:\!\!-\:\!\!\bar{Z})
      ( \partial^\mu \varphi_\fla^*\;\!V_{\fla \mu}\!
      + V^*_{\fla \mu} \partial^\mu \varphi_\fla ) \Big] \nn \\* &
    - \frac{i\kappa^2\!}{4}\,m_\fla\:\!
      (\varphi_\fla^* V_{\fla \mu}\!+ V^*_{\fla \mu} \varphi_\fla)
      (\bar{Z} \partial^\mu\:\!\!Z\:\!\!-\:\!\!Z \partial^\mu\:\!\!\bar{Z} )
    + \frac{\kappa^2\!}{8}
      \Big[ \varphi_\fla^* \varphi_\fla \partial_\mu
            \bar{Z}\;\!\partial^\mu Z\:\!\!
          +\:\!\!\bar{Z} Z \partial_\mu \varphi_\fla^*\;\!
            \partial^\mu \varphi_\fla
      \Big]
      \bigg\} \nn \\*
    + \sum_{\!\fla, \flb=1\!\!}^{N_f} \bigg\{ &
      \frac{\kappa^2\!}{8} \varphi_\fla^* \varphi_\fla
      \Big[ \partial_\mu \varphi_\flb^*\;\!\partial^\mu \varphi_\flb
      - 3 m_\flb^2 \varphi_\flb^* \varphi_\flb
      + 2 m_\flb^2 V_{\flb \mu}^* V_\flb^\mu \Big] \nn \\* &
    + \frac{\kappa^2\!}{4} m_\fla m_\flb
      \Big[ \varphi_\fla^* \varphi_\flb^* V_{\fla \mu} V_\flb^\mu
      + \varphi_\fla \varphi_\flb V_{\fla \mu}^* V_\flb^{*\mu}
      + 2\varphi_\fla^* \varphi_\flb V_{\flb \mu}^* V_\fla^\mu \Big]
      \bigg\} + {\cal O}(\kappa^3) \, .
\label{PertLagrangian}
\end{align}}%
\normalsize
Here $R$ is the Ricci scalar curvature and $Z$ is the complex axiodilaton scalar field. The massive matter fields are denoted according to \tab{tab:particlecontent}, and $\fla$, $\flb$ are their flavor indices.  We follow the notation in which Lagrangians and actions are related as $S=\!\int\!d^4x \sqrt{-g} {\cal L}$.
We have explicitly checked through order $\kappa^2$
that the Lagrangian~\eqref{PertLagrangian} is consistent with the amplitudes given by the double copy of QCD.

\section{Double copy in free theory}
\label{sec:wavefunctions}

In this section we review the external wavefunctions
for massive particles using the massive spinor-helicity formalism
of \rcite{Arkani-Hamed:2017jhn}.\footnote{The specific spinor-helicity conventions
we use here are detailed in the latest arXiv version of \rcite{Ochirov:2018uyq}.}
As in the purely massless case,
the idea of this formalism is to build up all of the scattering kinematics
from basic ${\rm SL}(2,\mathbb{C})$ spinors
such that all resulting formulae are covariant
with respect to the little group of the each physical particle.
For massive particles in four dimensions the little group is ${\rm SU}(2)$.
As usual, massive momenta are constructed from two-spinors
using the Pauli matrices:
\be
    p_{\alpha\dot{\beta}} = p_{\mu} \sigma^\mu_{\alpha\dot{\beta}}
    = \epsilon_{ab} \ket{p^a}_\alpha [p^b|_{\dot{\beta}}
    = \ket{p^a}_\alpha  [p_a|_{\dot{\beta}}
    = \la_{\alpha}^{~a} \lb_{\dot{\beta}a} ,
\label{eq:massivespinors}
\ee
where the ${\rm SU}(2)$ little-group indices $a,b,\ldots=1,2$
should be distinguished from the spinorial ${\rm SL}(2,\mathbb{C})$ indices
$\alpha,\beta,\ldots=1,2$ and $\dot{\alpha},\dot{\beta},\ldots=1,2$,
which represent the Lorentz group.
In that sense, the latter indices can be thought of as off-shell,
whereas the little-group indices encode the on-shell spin degrees of freedom.
In fact, any scattering amplitude can be represented~\cite{Arkani-Hamed:2017jhn}
as having $2s$ symmetric indices for each $s$-spin massive state with momentum $p$,
as carried by the helicity spinors
$\la_{p\:\!\alpha}^{~\;a}$ and~$\lb_{p}^{\dot{\alpha}a}$.

\subsection{Free spin 1/2}
\label{sec:spin1half}

The most familiar massive particle with spin is the Dirac fermion.
In the Weyl basis of the Clifford algebra,
its external Dirac spinors can be built from the helicity two-spinors as
\begin{subequations} \begin{align}
   u_{p}^{a} &
    = \begin{pmatrix} \la_{p\:\!\alpha}^{~\;a} \\ \lb_{p}^{\dot{\alpha}a}
      \end{pmatrix} , \qquad \quad
   \bar{u}_{p}^{a}
    = \begin{pmatrix} -\la_{p}^{\alpha a} \\ ~~\lb_{p\:\!\dot{\alpha}}^{~\;a}
      \end{pmatrix}
   \quad\;\Rightarrow \qquad
   \left\{
   \begin{aligned}
    & (\!\not{\!p}-m) u_{p}^{a} = \bar{u}_{p}^{a} (\!\not{\!p}-m) = 0 \, , \\
    & u_{p}^a \bar{u}_{p\:\!a} = u_{p}^{a} \epsilon_{ab} \bar{u}_{p}^{b}
    = \not{\!p}+m\, ,
   \end{aligned}
   \right.
\label{diracmassiveu} \\
   v_{p}^{a} &
    = \begin{pmatrix}-\la_{p\:\!\alpha}^{~\;a} \\ ~~\lb_{p}^{\dot{\alpha}a}
      \end{pmatrix} \, , \qquad\;
   \bar{v}_{p}^{a}
    = \begin{pmatrix} \la_{p}^{\alpha a} \\ \lb_{p\:\!\dot{\alpha}}^{~\;a}
      \end{pmatrix}
   \qquad \Rightarrow \qquad
   \left\{
   \begin{aligned}
    & (\!\not{\!p}+m) v_{p}^{a} = \bar{v}_{p}^{a} (\!\not{\!p}+m) = 0 \, , \\
    & v_{p}^a \bar{v}_{p\:\!a} = v_{p}^{a} \epsilon_{ab} \bar{v}_{p}^{b}
    = -\!\!\not{\!p}+m \, ,
   \end{aligned}
   \right.
\label{diracmassivev}
\end{align} \label{diracmassiveuv}%
\end{subequations}
as already explored in \rcites{Arkani-Hamed:2017jhn,Ochirov:2018uyq}.
We follow the textbook convention in that we normalize these spinors to $2m$,
\be
   \bar{u}_{p\:\!a} u_{p}^{b} = \bar{v}_{p\:\!a} v_{p}^{b} = 2m \delta_a^b\, .
\label{diracmassivenorm}
\ee
This identity illustrates the general fact
that the upper and lower little-group indices are related by complex conjugation,
which will also be evident from numerous equations below.

Let us use the Pauli-Lubanski (pseudo)vector to consider spin.
More precisely, for $m \neq 0$ we use its mass-rescaled version
as the spin vector
\be
   S_\lambda = \frac{1}{2m} \epsilon_{\lambda\mu\nu\rho} S^{\mu\nu} p^\rho ,
\label{paulilubanski}
\ee
where $S^{\mu\nu}$ is the intrinsic angular momentum operator that satisfies
the Lorentz algebra
\be
   \big[ S^{\lambda\mu},S^{\nu\rho} \big] =
     - i \eta^{\lambda\nu} S^{\mu\rho}
     + i \eta^{\lambda\rho} S^{\mu\nu}
     + i \eta^{\mu\nu} S^{\lambda\rho}
     - i \eta^{\mu\rho} S^{\lambda\nu} .
\label{lorentzalgebra}
\ee
In the Dirac case, it is explicitly
\be
   S^{\mu\nu}_{s=1/2} 
    = \frac{i}{4} [\gamma^\mu,\gamma^\nu]
    = \begin{pmatrix}
          \sigma_{~~~\alpha}^{\mu\nu,~\,\beta}\! & 0 \\
          0 & \bar{\sigma}^{\mu\nu,\dot{\alpha}}_{~~~~\,\dot{\beta}}
      \end{pmatrix} ,
\label{LorentzGeneratorsDirac}
\ee
where $\sigma_{~~~\alpha}^{\mu\nu,~\,\beta}$
and $\bar{\sigma}^{\mu\nu,\dot{\alpha}}_{~~~~\,\dot{\beta}}$
form separate left- and right-handed representations
of the Lorentz algebra~\eqref{lorentzalgebra},
that are selfdual and anti-selfdual, respectively.
Rewriting the spin operator as
$S^\mu = -\frac{1}{4m} [\gamma^\mu,\!\not{\!p}\,] \gamma^5$,
we can directly compute its one-particle matrix element
\be
   \bar{u}_{p}^{a}\:\!S^\mu u_p^{b}
    =-\frac{1}{2}
      \big( \bra{p^a}\sigma^\mu|p^b] + [p^a|\bar{\sigma}^\mu\ket{p^b} \big)\,,
\label{spinhalf}
\ee
and observe that it gives the exact same value
as the familiar Dirac-spin operator $\gamma^\mu \gamma^5/2$.
We will rely on the associated textbook spin vector
$s^\mu(u_s) = \frac{1}{2m} \bar{u}_s \gamma^\mu \gamma^5 u_s$ ---
only we rewrite it using the little-group indices
\be
   s_p^\mu = \frac{1}{2m}\:\!\bar{u}_{p\:\!1}\;\!\gamma^\mu \gamma^5 u_p^1
           =-\frac{1}{2m}\:\!\bar{u}_{p\:\!2}\;\!\gamma^\mu \gamma^5 u_p^2
   \qquad \Rightarrow \qquad
   s_p^2 = -1 \, , \qquad p \cdot s_p=0 \, .
\label{unitspinvector}
\ee
Here we have used that the two spin states of the Dirac fermion
have opposite spin vectors, and picked one of them.
Now recall that the little-group indices $a=1,2$
permit ${\rm SO}(3)$ rotations of the spin quantization axis.
This axis can be set to the three-momentum of the particle,
which corresponds to the definite-helicity spinor parametrization
detailed in \rcites{Arkani-Hamed:2017jhn,Ochirov:2018uyq}.
For a momentum parametrized as
$p^\mu=(E,P\cos\varphi \sin\theta,P\sin\varphi \sin\theta,P\cos\theta)$,
the definite-helicity spin vector is explicitly
\beal
   s_p^\mu = \frac{1}{m}
      (P,E\cos\varphi \sin\theta,E\sin\varphi \sin\theta,E\cos\theta) \, .
\label{massivespinvector}
\eeal
However, even in a generic representation
the one-particle expectation value of the spin operator is
\be
   \braket{S^\mu}_p^a
    = \frac{ \bar{u}_{p\:\!a} S^{\mu} u_p^{a} }
           { \bar{u}_{p\:\!a} u_p^{a} }
    = \left\{
      \begin{aligned}
      s_p^\mu/2 &, \quad a=1 \,, \\
     -s_p^\mu/2 &, \quad a=2 \,.
      \end{aligned} \right.
\ee

Our conjugation conventions for the external spinors are
\beal
   \big( \la_{p\:\!\alpha}^{~\;a} \big)^* &
    = \sgn(p^0) \lb_{p\:\!\dot{\alpha}\:\!a}\,  , \qquad\:\!\!\quad
   (u^a_p)^\dagger = \sgn(p^0)\bar{u}_{p\:\!a} \gamma^0 , \qquad\:\!\!\quad
   v_{p\:\!a} = \sgn(p^0) C (u_p^a)^* \, , \\
   \big( \lb_{p\:\!\dot{\alpha}}^{~\;a} \big)^* &
    =-\sgn(p^0) \la_{p\:\!\alpha\:\!a} \, , \qquad
   (v^a_p)^\dagger = -\sgn(p^0)\bar{v}_{p\:\!a} \gamma^0 , \qquad
   u_{p\:\!a} = -\sgn(p^0) C (v_p^a)^* \, ,
\label{diracconjugation}
\eeal
consistent with the massless convention
$(\la_{p\:\!\alpha})^*= \sgn(p^0) \lb_{p\:\!\dot{\alpha}}$.
Here we have used the charge-conjugation operator $C$, which plays a crucial role
when one wishes to impose the Majorana reality condition on the fermionic fields
\be
   \Psi_{\rm M} = \Psi_{\rm M}^{\rm c} = C \Psi^* , \qquad \quad
   C = C^\mathsf{T} = C^\dagger = C^{-1}
    = \begin{pmatrix}
         0 & \epsilon_{\alpha\beta} \\
         \epsilon^{\dot{\alpha}\dot{\beta}} & 0
      \end{pmatrix} .
\label{MajoranaRealityCondition}
\ee
As detailed in \app{app:majorana},
the properties~\eqref{diracconjugation}
imply that the same external spinors~\eqref{diracmassiveuv}
may be used to construct amplitudes with external Majorana fermions.
The translation from QCD to its self-conjugate version~\cite{Anber:2019nfu}
is hence straightforward.

\subsection{Free spin 1}
\label{sec:spin1}

Spins higher than one are represented
by $2s$ symmetrized little-group indices~\cite{Arkani-Hamed:2017jhn}.
The effect of this symmetrization first shows itself for massive vector particles.
Their polarization vectors
\be
   \varepsilon_{p\:\!\mu}^{ab}
    = \frac{i \bra{p^{(a}}\sigma_\mu|p^{b)}]}{\sqrt{2}m}
   \qquad \Rightarrow \qquad
   \left\{
   \begin{aligned}
   p\cdot\varepsilon_{p}^{ab} & = 0 \, , \\
   \varepsilon_{p}^{ab} \cdot \varepsilon_{p\:\!cd} &
    = - \delta^{(a}_{(c} \delta^{b)}_{d)} \, , \\
   \varepsilon_{p\:\!\mu}^{ab} \varepsilon_{p\:\!\nu\:\!ab} &
    = - \eta_{\mu\nu} + \frac{p_\mu p_\nu}{m^2} \, , \\
   (\varepsilon_{p\:\!\mu}^{ab})^* & = \varepsilon_{p\:\!\mu\:\!ab}
    = \epsilon_{ac} \epsilon_{bd} \varepsilon_{p\:\!\mu}^{cd} \, ,
   \end{aligned}
   \right.
\label{polvectorsmassive}
\ee
were first spelled out in \rcites{Guevara:2018wpp,Chung:2018kqs}.
They are transverse, as required by the Proca equation:
\be
   \begin{aligned}
   {\cal L}_\text{Proca} &
     =-\frac{1}{2} V_{\mu\nu}^* V^{\mu\nu} + m^2 V_\mu^* V^\mu \\ &
   \Rightarrow \qquad
   -\partial_\mu V^{\mu\nu} = m^2 V^\nu
   \end{aligned}
   \qquad \Rightarrow \qquad
   \left\{
   \begin{aligned}
      (\partial^2+m^2) V^\nu & = 0 \, , \\
      \partial_\mu V^\mu & = 0 \, ,
   \end{aligned}
   \right.
\label{ProcaLagrangian}
\ee
where $V_{\mu\nu} = \partial_\mu V_\nu - \partial_\nu V_\mu$
is the field strength of the massive vector.
To see how the spin is encoded by the symmetric indices $a,b=1$,
we use the Lorentz generators in the vector representation
\be
   (S^{\mu\nu}_{s=1})^{\rho}_{~\,\sigma}
    = i (\eta^{\mu\rho} \delta^\nu_\sigma - \eta^{\nu\rho} \delta^\mu_\sigma) \, .
\label{LorentzGeneratorsVector}
\ee
They feed into the Pauli-Lubanski vector~\eqref{paulilubanski},
and its one-particle expectation value is explicitly
quantized in terms of the unit spin vector~\eqref{unitspinvector}:
\be
   \braket{S^\mu}_p^{ab}
    = \frac{ \varepsilon_{p\:\!\rho\:\!ab}
             S^{\mu,\rho}_{~~~~\sigma} \varepsilon_p^{\sigma\:\!ab} }
           { \varepsilon_{p\:\!ab} \cdot \varepsilon_p^{ab} }
    = \left\{
      \begin{aligned}
      s_p^\mu &, \quad a=b=1\, , \\
      0 &, \quad a+b=3 \,, \\
     -s_p^\mu &, \quad a=b=2 \,,
      \end{aligned} \right.
\ee
Notice that the spacelike normalization in \eqn{polvectorsmassive} implies
$ \varepsilon_{p\:\!11}\!\cdot\!\varepsilon_{p}^{11}
= \varepsilon_{p\:\!22}\!\cdot\!\varepsilon_{p}^{22} = -1$
but at the same time $ \varepsilon_{p\:\!12}\!\cdot\!\varepsilon_{p}^{12} = -1/2$.
However, this turns out to be convenient:
in the completeness relation, for instance,
the factor of $1/2$ is automatically compensated
by summing over $\varepsilon_{p}^{12}$ and its equal $\varepsilon_{p}^{21}$.
One can also confirm that in the massless limit
the longitudinal polarization vector $\varepsilon_{p}^{\mu\:\!12}$
behaves as $p^\mu/m$, up to a numerical factor.

There is a hint at the massive double-copy relation
already in the above definition of the spin-1 wavefunction,
which we can rewrite as a bi-spinor
\be
   \varepsilon_{p\:\!\alpha\dot{\beta}}^{ab}
    = \frac{i\sqrt{2}}{m} \la_{p\:\!\alpha}^{~(a} \lb_{p\:\!\dot{\beta}}^{~\;b)} \,.
\ee
This is reminiscent of the well-known fact
that a symmetric tensor product of two spinors gives a vector.
The remaining antisymmetric combination of spinors can be regarded as a scalar
via the spinor identities
\be
   \la_{p\:\!\alpha}^{~[a} \lb_{p\:\!\dot{\beta}}^{~\;b]}
    = -\frac{1}{2} p_{\alpha\dot{\beta}} \epsilon^{ab} \,, \qquad \quad
   \la_{p\:\!\alpha}^{~[a} \la_{p\:\!\beta}^{~\;b]}
    = -\frac{m}{2} \epsilon_{\alpha\beta} \epsilon^{ab} \,, \qquad \quad
   \lb_{p\:\!\dot{\alpha}}^{~[a} \lb_{p\:\!\dot{\beta}}^{~\;b]}
    = -\frac{m}{2} \epsilon_{\dot{\alpha}\dot{\beta}} \epsilon^{ab} \, .
\ee
Of course, it is difficult to interpret the above equations as strict double copies
at the level of external wavefunctions.
For instance, the anti-aligned Dirac arrows can be shown to give
rather perplexing identities:
\be
   u_p^{(a} \bar{v}_p^{b)} = -\frac{i}{\sqrt{2}}
      ( \gamma^0\!\!\not{\!p}^{\dagger}\gamma^0 + m )\!\not{\!\varepsilon}_p^{ab} \,,
   \qquad \quad
   v_p^{(a} \bar{u}_p^{b)} = -\frac{i}{\sqrt{2}}
      ( \gamma^0\!\!\not{\!p}^{\dagger}\gamma^0 - m )\!\not{\!\varepsilon}_p^{ab} \,.
\ee
In \sec{sec:3pt}, however, we will show
that the symmetric and antisymmetric combinations of QCD amplitudes
with Dirac or Majorana spinors combine exactly into gravitational amplitudes
with massive vector and scalar bosons, respectively.

\subsection{Higher spins}
\label{sec:spin3halves}

Here let us remark here that the external wavefunctions for spins higher than one
are built out of the lower-spin ones
in a way that explicitly respects the double-copy construction.
The integer-spin polarization tensors are simple symmetric tensor products
of the vectors~\eqref{polvectorsmassive}~\cite{Guevara:2018wpp,Chung:2018kqs}
\be
   \varepsilon_{p\:\!\mu_1\ldots\:\!\mu_s}^{a_1\ldots\:\!a_{2s}}
    = \varepsilon_{p\:\!\mu_1}^{(a_1 a_2}\ldots\:\!
      \varepsilon_{p\:\!\mu_s}^{a_{2s-1} a_{2s})} \,.
\label{poltensors}
\ee
The construction is similar for half-integer spins,
and it may be instructive to consider massive Rarita-Schwinger particles
here in more detail.

The external wavefunctions for spin-3/2 particles
considered \eg in \rcites{Zinoviev:2007ig,Freedman:2012zz}
can be constructed as explicit double copies those for spin 1/2 and 1:
\be
   \begin{aligned}
   u_{p\:\!\mu}^{abc} & = u_p^{(a} \varepsilon_{p\:\!\mu}^{bc)} \\
   \bar{u}_{p\:\!\mu}^{abc} & = \varepsilon_{p\:\!\mu}^{(ab} \bar{u}_p^{c)}
   \end{aligned}
   \qquad \Rightarrow \qquad\!\!\!
   \left\{
   \begin{aligned}
    & (\gamma^{[\la\mu\nu]} p_\mu  + m \gamma^{[\la\nu]}) u_{p\:\!\nu}^{abc}
    = \bar{u}_{p\:\!\la}^{abc} (\gamma^{[\la\mu\nu]} p_\mu + m \gamma^{[\la\nu]})
    = 0 \, , \\
    & p^\mu u_{p\:\!\mu}^{abc} = \bar{u}_{p\:\!\mu}^{abc} p^\mu = 0\,  , \\
    & \gamma^\mu u_{p\:\!\mu}^{abc} = \bar{u}_{p\:\!\mu}^{abc} \gamma^\mu = 0 \, , \\
   \end{aligned}
   \right.
\label{polspinors}
\ee
and similarly for the $v_{p\:\!\mu}^{abc}$.
The angular momentum operator is constructed accordingly
as a direct sum of the spin-1/2 and spin-1 ones:
\be
   (S^{\mu\nu}_{s=3/2})^{A\rho}_{~~\;B\sigma}
    = (S^{\mu\nu}_{s=1/2})^{A}_{~\:B}\;\!\delta^{\rho}_\sigma
    + \delta^{A}_{B} (S^{\mu\nu}_{s=1})^{\rho}_{~\,\sigma} \, .
\label{LorentzGeneratorsRS}
\ee
The expectation value of the Pauli-Lubanski operator gives
the expected four states of the massive Rarita-Schwinger particle:
\vspace{-10pt}
\be
   \braket{S^\mu}_p^{abc}
    = \frac{ \bar{u}_{p\:\!\rho\:\!abc} S^{\mu,\rho}_{~~~~\;\sigma}
             u_p^{\sigma\:\!abc} }
           { \bar{u}_{p\:\!\sigma\:\!abc} u_p^{\sigma\:\!abc} }
    = \left\{
      \begin{aligned}
      3 s_p^\mu/2 &, \quad a=b=c=1 \,, \\
      s_p^\mu/2 &, \quad a+b+c=4 \,, \\
     -s_p^\mu/2 &, \quad a+b+c=5 \,, \\
     -3 s_p^\mu/2 &, \quad a=b=c=2 \,.
      \end{aligned} \right.
\ee
Similarly to Dirac fermions,
we normalized the wavefunctions using the particle's mass:
\be
   \bar{u}_{p\:\!111}\!\cdot\!u_{p}^{111}
    = \bar{u}_{p\:\!222}\!\cdot\!u_{p}^{222}
    = 3 \bar{u}_{p\:\!112}\!\cdot\!u_{p}^{112}
    = 3 \bar{u}_{p\:\!122}\!\cdot\!u_{p}^{122} = -2m \, ,
\ee
where the sign is due to them being spacelike in the mostly-minus metric convention.
Again, the normalization of the lower-helicity states,
such as $\bar{u}_{p\:\!112}=\bar{u}_{p\:\!121}=\bar{u}_{p\:\!211}$,
cancels their overcounting in state sums of the form
$u_{p\:\!\mu}^{abc} \bar{u}_{p\:\!\nu\:\!abc}$ .

The formulae given here demonstrate that the massive-spinor helicity formalism
is very useful if one wishes
to consider a quantum field theory of higher-spin states.
In this paper, however, we restrict ourselves to examples
with lower-spin matter content, as present in QCD ($s \le 1/2$) and its double copy ($s \le 1$).

\section{Double copy in three-point vertices}
\label{sec:3pt}

In this section we explore the double copy
of the three-point matter amplitudes in QCD.

Note that at three points there is no need to distinguish between
the KLT~\cite{Kawai:1985xq} and the BCJ double copy~\cite{Bern:2008qj},
as the full amplitude consists of one gauge-invariant diagram,
and both constructions amount to
\be
   {\cal M}_3 = \frac{i\kappa}{2} A_3 \, \widetilde A_3 \, ,
\label{dc3pt}
\ee
where $A_3$ and $\widetilde A_3 $ are  color-stripped coupling-stripped gauge-theory amplitudes.
For instance, the gluonic self-interaction vertex amplitudes\footnote{In this paper
we use the amplitude-friendly conventions where all momenta are taken incoming
and the color generators are normalized to satisfy
$\Tr(T^{a} T^{b})=\delta^{ab}$ and $[T^a,T^b]=\tf^{abc}T^c$.
}
\be
   {\cal A}(1^+_a\:\!\!,2^-_b\:\!\!,3^-_c) = -ig\tilde{f}^{abc}
      \frac{\braket{2\;\!3}^3}{\braket{1\;\!2} \braket{3\;\!1}} \,, \qquad \quad
   {\cal A}(1^-_a\:\!\!,2^+_b\:\!\!,3^+_c) = ig\tilde{f}^{abc}
      \frac{[2\;\!3]^3}{[1\;\!2] [3\;\!1]} \, ,
\label{gluon3pt}
\ee
double-copy to the gravitational vertices (with $g \to \kappa/2$)
\be
   {\cal M}(1^+\:\!\!,2^-\:\!\!,3^-) = -\frac{i\kappa}{2}
      \frac{\braket{2\;\!3}^6}{\braket{1\;\!2}^2 \braket{3\;\!1}^2} \,, \qquad \quad
   {\cal M}(1^-\:\!\!,2^+\:\!\!,3^+) = -\frac{i\kappa}{2}
      \frac{[2\;\!3]^6}{[1\;\!2]^2 [3\;\!1]^2} \,.
\label{graviton3pt}
\ee

The massive-matter results, summarized in \sec{sec:doublecopy3pt},
are very simple and, up to minor modifications,
satisfy the double-copy relation~\eqref{dc3pt}.
For completeness, we go through the cases of $s=0$, $1/2$ and $1$
both in gauge theory and in gravity and compute the three-point amplitudes
from explicit Feynman rules.
For consistency purposes,
we consider complex matter
in a fundamental representation ---
without relying on specific details of the gauge group.
This means that the representation can be easily switched, for example,
to a real representation.
Then the exact same Feynman rules as given below for complex matter
may be used for self-conjugate matter,
the least trivial case being the switch between Dirac and Majorana fermions,
as reviewed in \app{app:majorana}.
As usual in that case, the necessary factors of $1/2$ in the action
are then compensated by the combinatoric factors
due to self-conjugacy.

\subsection{Three-vertex for spin 0}
\label{sec:v3spin0}

In this section we warm up with a scalar in gauge theory and gravity.
The familiar Lagrangian
\be
   {\cal L}_\text{scalar} = g^{\mu\nu} (D_\mu \varphi)^\dagger (D_\nu \varphi)
    - m^2 \varphi^\dagger \varphi
\ee
implements both couplings minimally.
Taking $g^{\mu\nu}=\eta^{\mu\nu}$ and
$D_\mu = \partial_\mu - i g A_\mu$
gives the gauge-theory vertex
\be
\label{GaugeVertexScalar}
   {\cal L}_{\varphi \varphi A} = i g
      \big[ \varphi^\dagger A^\mu (\partial_\mu \varphi)
          - (\partial_\mu \varphi^\dagger) A^\mu \varphi
      \big]
   \qquad \Rightarrow \quad~
   \scalegraph{0.9}{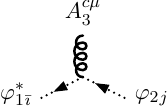}\!=\frac{ig}{\sqrt{2}} T_{i \bar\jmath}^c (p_2-p_1)^\mu ,
\ee
where all the momenta are taken to be incoming.
On the other hand, considering the massive scalar to be colorless
and expanding the metric
$\sqrt{-g} g^{\mu\nu} = \eta^{\mu\nu} - \kappa \hh^{\mu\nu}$,
we find the gravitational coupling vertex
\be
\label{GravityVertexScalar}
   (\sqrt{-g} {\cal L})_{\varphi \varphi \hh}
    = -\kappa \hh^{\mu\nu} \partial_\mu \varphi^*\;\!\partial_\nu \varphi
    + \frac{\kappa}{2} m^2 \varphi^* \varphi\;\!\hh
   \qquad \Rightarrow \quad
   \scalegraph{0.9}{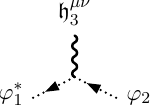}\!= i\kappa \bigg[ p_1^{(\mu} p_2^{\nu)}
    + \frac{m^2\!}{2} \eta^{\mu\nu} \bigg] ,
\ee
where the second contribution in the bracket is a trace term due to
$\sqrt{-g} = 1 - \kappa \hh/2 + O(\kappa^2)$ from the spacetime measure.
The corresponding three-point amplitudes are
\be
   {\cal A}(1_i,2_{\bar \jmath},3_c)
    = -ig T_{i \bar\jmath}^c \sqrt{2} (p_1\cdot\varepsilon_3)\, , \qquad \quad
   {\cal M}(1,2,3)
    = -i\kappa (p_1\cdot\varepsilon_3)^2 ,
\label{ScalarMatter3pt}
\ee
where we used momentum conservation $p_1+p_2+p_3=0$
for all states considered incoming,
as well as the transverse and traceless polarization tensor
$\varepsilon_3^{\mu\nu} = \varepsilon_3^\mu \varepsilon_3^\nu$ of the graviton.
The double-copy relation~\eqref{dc3pt} is now evident.

This scalar example is convenient for explaining the helicity variable $x_3$,
first introduced in \rcite{Arkani-Hamed:2017jhn}.
It follows from the on-shell conditions for the three-point kinematics
that there exists a proportionality coefficient~$x_3$
between the spinors $\ket{3}$ and $|1|3]$:
\be
   p_1^2-m^2 = \bra{3}2|3] = 0 \, , \qquad \quad
   p_2^2-m^2 = \bra{3}1|3] = 0 \qquad \Rightarrow \qquad
   |1|3] = - |2|3] \equiv -m x_3 \ket{3}\, .
\ee
Therefore, we have
\be
   x_3 = \frac{[3|1\ket{q}}{m \braket{3\;\!q}}
       = -\sqrt{2}\:\!\frac{p_1\cdot\varepsilon_3^+\!}{m} \,, \qquad \quad
   x_3^{-1} = \frac{\bra{3}1|q]}{m [3\;\!q]}
       =  \sqrt{2}\:\!\frac{p_1\cdot\varepsilon_3^-\!}{m} \,.
\label{xFactor}
\ee
where we have used an arbirary pair of spinors $\ket{q}$ and $|q]$ to write
an explicit but nonlocal expression for $x_3$. It is, however,
independent of them as long as $\braket{3\,q} \neq 0 \neq [3\,q]$.
The coefficient $x_3$ is designed to be dimensionless
and carry a unit positive helicity with respect to momentum $p_3$.
The three-point amplitudes~\eqref{ScalarMatter3pt}
are then proportional to the appropriate powers of this helicity factor:
\begin{subequations} \begin{align}
   {\cal A}(1_i\:\!\!,2_{\bar \jmath},3^+_c) &
    = i g T_{i \bar\jmath}^c m x_3 \,, \qquad\:\,\qquad
   {\cal M}(1,2,3^+) = -\frac{i\kappa}{2} m^2 x_3^2 \,, \\*
   {\cal A}(1_i\:\!\!,2_{\bar \jmath},3^-_c) &
    =-i g T_{i \bar\jmath}^c m x_3^{-1} , \qquad \quad
   {\cal M}(1,2,3^-) = -\frac{i\kappa}{2} m^2 x_3^{-2} \, .
\end{align} \label{ScalarMatter3pt2}%
\end{subequations}

\subsection{Three-vertex for spin 1/2}
\label{sec:v3spin1half}

In this section we consider a spin-1/2 particle
minimally coupled to a gauge or gravitational field.
For simplicity, we consider the complex case of the Dirac fermion
\be
\label{DiracLagrangian}
   {\cal L}_\text{Dirac} = \bar{\Psi} (i\gamma^\mu D_\mu - m) \Psi \,,
\ee
with the projection to the Majorana case expained in \app{app:majorana}.

In gauge theory, the three-point vertex is evidently
\be
\label{GaugeVertexDirac}
   {\cal L}_{\bar{\Psi} \Psi A} = g \bar{\Psi} \!\not{\!\!A} \Psi 
   \qquad \Rightarrow \quad~
   \scalegraph{0.9}{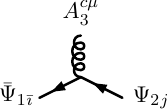}\!= \frac{i g}{\sqrt{2}} T_{i \bar\jmath}^c \gamma^\mu .
\ee
We dress it with the external spinors $\bar{v}_1^a$ and $u_2^b$
from \eqn{diracmassiveuv} and use a Schouten identity
to write the resulting amplitudes in the spinor-helicity form:
\begin{subequations} \begin{align}
\label{FermionMatter3ptGauge1}
   {\cal A}(1^a_i\:\!\!,2^b_{\bar \jmath},3^+_c) &
    = \frac{i g T_{i \bar\jmath}^c}{\braket{q\;\!3}}
      \big( \braket{1^a q} [3\;\!2^b] + [1^a 3] \braket{q\;\!2^b} \big)
    = i g T_{i \bar\jmath}^c \braket{1^a 2^b} x_3 \,, \\*
\label{FermionMatter3ptGauge2}
   {\cal A}(1^a_i\:\!\!,2^b_{\bar \jmath},3^-_c) &
    =-\frac{i g T_{i \bar\jmath}^c}{[q\;\!3]}
      \big( \braket{1^a 3} [q\;\!2^b] + [1^a q] \braket{3\;\!2^b} \big)
    =-i g T_{i \bar\jmath}^c [1^a 2^b] x_3^{-1} .
\end{align} \label{FermionMatter3ptGauge}%
\end{subequations}

Extracting the gravitational three-point vertex is somewhat more involved (its derivation from the Noether energy-momentum tensor is given in \app{app:gravitydirac}). It gives
\be
\label{GravityVertexDirac}
   \scalegraph{0.9}{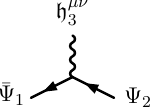}\!= \frac{i\kappa}{4} (p_1-p_2)^{(\mu} \gamma^{\nu)}
    + \text{off-shell/trace terms} .
\ee
Plugging in the spinors and the polarization tensor is then
no more challenging than in the case of QCD.
We immediately get the amplitudes
\be
   {\cal M}(1^a,2^b,3^+) = -\frac{i \kappa}{2} m \braket{1^a 2^b} x_3^2 \, ,
   \qquad \quad 
   {\cal M}(1^a,2^b,3^-) = -\frac{i \kappa}{2} m [1^a 2^b] x_3^{-2} \, .
\label{FermionMatter3ptGravity}
\ee
They obey a simple massive implementation of the double-copy relation~\eqref{dc3pt}
in the sense of multiplying the spin-1/2 amplitudes~\eqref{FermionMatter3ptGauge}
with their scalar counterparts in \eqn{ScalarMatter3pt2}.
Such gravitationally-interacting fermionic matter is, however,
not part of the double copy of standard QCD, but it is relevant for supersymmetric versions of QCD, or for cases when some quarks are replaced by fundamental scalars.

\subsection{Three-vertex for spin 1}
\label{sec:v3spin1}

In gauge theory, the interaction vertex for a massive vector $V_\mu$ obtained from spontaneous symmetry breaking, is given by
(see \eg \rcite{Peskin:1995ev})
\begin{align}
\label{GaugeVertexVector}
   {\cal L}_{VV\!A} & = -ig
      \big[ V_{\mu\nu}^\dagger A^\nu V^\mu - V_\mu^\dagger A_\nu V^{\mu\nu}
           + V_\mu^\dagger F^{\mu\nu} V_\nu
      \big] \\* & \Rightarrow~
   \scalegraph{0.9}{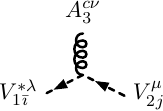}\!
    = \frac{ig}{\sqrt{2}} T_{i \bar\jmath}^c
      \big[ \eta^{\la\mu} (p_1-p_2)^\nu
          + \eta^{\mu\nu} (p_2-p_3)^\lambda
          + \eta^{\nu\la} (p_3-p_1)^\mu \big] \, . \nn
\end{align}
Note that it does not coincide with what one would obtain from
the free Proca Lagrangian~\eqref{ProcaLagrangian}
by a simple replacement $\partial_\mu \to D_\mu$~\cite{Chung:2018kqs}.
The non-abelian-like interaction is required for obtaining a scattering amplitude that is well-behaved in the massless limit~\cite{Arkani-Hamed:2017jhn}, which is compatible with spontaneous symmetry breaking.
Using the massive polarization vectors introduced in \eqn{polvectorsmassive}
and the three-point identities,
\be
   [1^a 3] = x_3 \braket{1^a 3}\, , \qquad \quad
   [2^b 3] = -x_3 \braket{2^b 3} \, , \qquad \quad
   [1^{a}2^{b}] = \braket{1^{a}2^{b}}
    - \frac{x_3}{m} \braket{1^a 3} \braket{3\;\!2^b}\, , 
\label{squaretoangle3pt2}
\ee
we are able to arrive at the following expressions
\begin{subequations} \begin{align}
   {\cal A}(1^{a_1 a_2}_i\:\!\!,2^{b_1 b_2}_{\bar \jmath}\:\!\!,3^+_c) &
    = -i g T_{i \bar\jmath}^c \frac{x_3}{m}
      \braket{1^{(a_1}\;\!2^{(b_1}} \braket{1^{a_2)}\:\!2^{b_2)}}
    = -i g T_{i \bar\jmath}^c \frac{\braket{1^a 2^b}^{\odot 2}\!}{m} x_3 \, , \\*
   {\cal A}(1^{a_1 a_2}_i\:\!\!,2^{b_1 b_2}_{\bar \jmath}\:\!\!,3^-_c) &
    = i g T_{i \bar\jmath}^c \frac{x_3^{-1}}{m}
      [1^{(a_1}\;\!2^{(b_1}]  [1^{a_2)}\:\!2^{b_2)}]
    = i g T_{i \bar\jmath}^c \frac{[1^a 2^b]^{\odot 2}\!}{m} x_3^{-1} .
\end{align} \label{VectorMatter3ptGauge}%
\end{subequations}
Here we have encoded
the symmetrization of the little-group indices for particles~1 and~2
using a symmetrized tensor product operation, denoted by~$\odot$.
It should be remembered that such a symmetrization is always applied to 
the spin indices of each massive particle independently.

While we do not include massive spin-1 bosons in the QCD Lagrangian, it is natural to consider double copies using spontaneously broken gauge theories, see \rcites{Chiodaroli:2015rdg,Chiodaroli:2017ehv,Chiodaroli:2018dbu}. For our purposes, however, the double copy of QCD does contain gravitating massive abelian vectors. 
Let us then consider a covariantization of the Proca action~\eqref{ProcaLagrangian},
which amounts to replacing the Lorentz contractions with
$\sqrt{-g} g^{\mu\nu} = \eta^{\mu\nu}\!-\kappa \hh^{\mu\nu}$.
Then the three-point vertex is
\begin{align}
\label{GravityVertexVector}
   (\sqrt{-g} {\cal L})_{V V \hh} &
    = \kappa \hh^{\mu\nu}
      \big( V^*_{\mu\sigma} V_{\nu}^{~\sigma} - m^2 V^*_\mu V_\nu \big)
    - \frac{\kappa}{4} \hh V^*_{\mu\nu} V^{\mu\nu} \\* &\;\!
   \begin{aligned}
       \Rightarrow~\scalegraph{0.9}{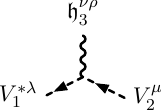} \\~
   \end{aligned}\!
   \begin{aligned}
    =-i\kappa
      \bigg[ & \big((p_1 \cdot p_2) + m^2\big) \eta^{\la(\nu} \eta^{\rho)\mu} \\ &~
           + \eta^{\la\mu} p_1^{(\nu} p_2^{\rho)}
           - p_1^\mu \eta^{\la(\nu} p_2^{\rho)} 
           - p_2^\la \eta^{\mu(\nu} p_1^{\rho)}  \\ & \qquad~~\,\quad
    - \frac{1}{2} \eta^{\nu\rho}
      \big( \eta^{\la\mu} (p_1 \cdot p_2) - p_2^\la p_1^\mu \big)
      \bigg] \, ,
   \end{aligned} \nn
\end{align}
where the last line is a trace term that
vanishes in the resulting three-point amplitude
\beal
   {\cal M}^{s=1}(\u{1},\o{2},3) =-i\kappa
      \big[ \big((p_1 \cdot p_2) + m^2\big)
            (\varepsilon_1\!\cdot\!\varepsilon_3)
            (\varepsilon_2\!\cdot\!\varepsilon_3) 
          + (\varepsilon_1\!\cdot\!\varepsilon_2)
            (p_1\!\cdot\!\varepsilon_3) (p_2\!\cdot\!\varepsilon_3) & \\
         -\,(\varepsilon_1\!\cdot\!\varepsilon_3)
            (p_2\!\cdot\!\varepsilon_3) (p_1\!\cdot\!\varepsilon_2)
          - (p_2\!\cdot\!\varepsilon_1) (p_1\!\cdot\!\varepsilon_3)
            (\varepsilon_2\!\cdot\!\varepsilon_3) &
      \big] \,.
\eeal
Plugging in the spinor-helicity variables, we find\footnote{In
\rcite{Guevara:2018wpp} the same amplitudes were derived
from the Noether energy-momentum tensor in flate space
symmetrized with a Belinfante-tensor contribution,
as we do here for the Dirac fermions in \app{app:gravitydirac}.
}
\be
   {\cal M}(1^{a_1 a_2}\:\!\!,2^{b_1 b_2}\:\!\!,3^+)
    = \frac{i\kappa}{2} \braket{1^a 2^b}^{\odot 2} x_3^2 \,, \qquad \quad
   {\cal M}(1^{a_1 a_2}\:\!\!,2^{b_1 b_2}\:\!\!,3^-) %
    = \frac{i\kappa}{2} [1^a 2^b]^{\odot 2} x_3^{-2} \,.
\label{VectorMatter3ptGravity}
\ee

\subsection{Three-point summary}
\label{sec:doublecopy3pt}

The three cases for $s=0$, $1/2$ and $1$ can be encapsulated
into the following formulae. The gauge-theoretic amplitudes
\begin{subequations} \begin{align}
   {\cal A}(1^{\{a\}}_i\:\!\!,2^{\{b\}}_{\bar \jmath}\:\!\!,3^{+}_c) &
    = (-1)^{\lfloor s \rfloor} ig\;\!T^c_{i\bar\jmath}
      \frac{\braket{1^a 2^b}^{\odot 2s}\!}{m^{2s-1}} x_3 \,, \\*
   {\cal A}(1^{\{a\}}_i\:\!\!,2^{\{b\}}_{\bar \jmath}\:\!\!,3^{-}_c) &
    = (-1)^{\lfloor s \rfloor+1}
      ig\;\!T^c_{i\bar\jmath} \frac{[1^a 2^b]^{\odot 2s}}{m^{2s-1}} x_3^{-1} ,
\end{align} \label{qcdmatter3pt}%
\end{subequations}
and their gravitational counterparts
\begin{subequations} \begin{align}
   {\cal M}(1^{\{a\}}\:\!\!,2^{\{b\}}\:\!\!,3^{+}) &
    = (-1)^{\lfloor s \rfloor+1} \frac{i\kappa}{2}
      \frac{\braket{1^a 2^b}^{\odot 2s}}{m^{2s-2}} x_3^{2} \,, \\*
   {\cal M}(1^{\{a\}}\:\!\!,2^{\{b\}}\:\!\!,3^{-}) &
    = (-1)^{\lfloor s \rfloor+1} \frac{i\kappa}{2}
      \frac{[1^a 2^b]^{\odot 2s}}{m^{2s-2}} x_3^{-2} \,.
\end{align} \label{gravitymatter3pt}%
\end{subequations}
are all consistent with the arbitrary-spin amplitudes
first proposed in \rcite{Arkani-Hamed:2017jhn}.
There, they were singled out of other possible spinor-helicity expressions
by their tame behavior in the massless limit,
which is said to define a notion of ``minimal coupling'' for massive matter
to the gauge and gravitational fields.
It now known to coincide with the textbook notion of minimal coupling
only for lower-spin cases~\cite{Chung:2018kqs},
on which we focus on in the present paper.
The purpose of the subtle sign prefactors in \eqns{qcdmatter3pt}{gravitymatter3pt}
is to match the Feynman-rule calculations in the mostly-minus metric convention:
they come from the standard requirement that the polarization tensors be spacelike.

It is evident that the arbitrary-spin amplitudes above obey
a simple extension of the double-copy relation~\eqref{dc3pt}
involving symmetrization of the little-group ${\rm SU}(2)$ indices,
in which gravitational matter of spin $s$ is constructed
from gauge-theoretic matter with spins~$s_1$ and~$s_2$:
\be\!\!
   {\cal M}(1^{\{a\}}_{s} \! ,2^{\{b\}}_{s}\! , 3)
    = (-1)^{\lfloor s \rfloor {-} \lfloor s_1 \rfloor {-} \lfloor s_2 \rfloor\,}
      \frac{i\kappa}{2}
      A(1^{\{a\}}_{s_1}\:\!\!,2^{\{b\}}_{s_1}\:\!\!,3) \odot
      A(1^{\{a\}}_{s_2}\:\!\!,2^{\{b\}}_{s_2}\:\!\!,3) \,, \qquad
   s = s_1 + s_2 \,.
\label{gravitymatter3ptKLT}
\ee
Below we consider in detail the three-point double copies
that can be constructed from the QCD vertices~\eqref{FermionMatter3ptGauge}.
This includes special cases of the little-group-symmetrized double copy above,
as well as less obvious little-group-antisymmetrized double copies.

\subsection{Double copy of QCD three-vertex}
\label{sec:qcd2gravity3pt}

As the simplest non-trivial example, let us consider
the double copy of the quark-gluon amplitudes~\eqref{FermionMatter3ptGauge}.
First, we consider the little-group-symmetrized combinations\footnote{We
indicate fields in the amplitude labels
when they are not evident from the context and the little-group labels.
In three-point amplitudes, only gravitational matter requires explicit labeling.
}
\begin{align}
   \frac{i\kappa}{2} A(1^{a_1}\:\!\!,2^{b_1}\:\!\!,3^+)\!\odot\!
      A(1^{a_2}\:\!\!,2^{b_2}\:\!\!,3^+) &
    =-\frac{i\kappa}{2}
      \braket{1^{(a_1}\;\!2^{(b_1}} \braket{1^{a_2)}\:\!2^{b_2)}} x_3^2
    =-{\cal M}(1^{a_1 a_2}\:\!\!,2^{b_1 b_2}\:\!\!,3^+) \, , \nn \\*
   \frac{i\kappa}{2} A(1^{a_1}\:\!\!,2^{b_1}\:\!\!,3^-)\!\odot\!
      A(1^{a_2}\:\!\!,2^{b_2}\:\!\!,3^-) &
    =-\frac{i\kappa}{2}
      [1^{(a_1}\;\!2^{(b_1}] [1^{a_2)}\:\!2^{b_2)}] x_3^{-2}\!
    =-{\cal M}(1^{a_1 a_2}\:\!\!,2^{b_1 b_2}\:\!\!,3^-) \, , \nn \\
\label{qcdDC3pt}
   \frac{i\kappa}{2} A(1^{a_1}\:\!\!,2^{b_1}\:\!\!,3^+)\!\odot\!
      A(1^{a_2}\:\!\!,2^{b_2}\:\!\!,3^-) & =
   \frac{i\kappa}{2} A(1^{a_1}\:\!\!,2^{b_1}\:\!\!,3^-)\!\odot\!
      A(1^{a_2}\:\!\!,2^{b_2}\:\!\!,3^+) \\*
    = \frac{i\kappa}{2}
      \braket{1^{(a_1}\;\!2^{(b_1}} [1^{a_2)}\:\!2^{b_2)}] &
    = \frac{i\kappa}{2} m^2 (\varepsilon_1^{a_1 a_2}\!\cdot\varepsilon_2^{b_1 b_2})
    =-{\cal M}(1^{a_1 a_2}\:\!\!,2^{b_1 b_2}\:\!\!,3_Z) \, . \nn
\end{align}
Here the first two amplitudes contain a graviton
and match the minimal-coupling formulae~\eqref{VectorMatter3ptGravity}.
The last two amplitude
is equal to $-i\kappa m^2 (\varepsilon_1\cdot\varepsilon_2)/2$ and
can be seen to correspond to a non-trivial coupling
of the axiodilaton scalar~$Z$ (or~$\bar{Z}$) to the vector mass term:
\beal
\label{GravityVertexVectorZ}
   {\cal L}_{VVZ} & = -\frac{\kappa}{2} (Z + \bar{Z}) m^2 V_\mu^* V^\mu
   \quad~\Rightarrow~~ 
   \scalegraph{0.9}{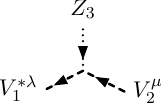}\!=\!\scalegraph{0.9}{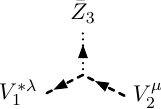}\!
    = -\frac{i\kappa}{2} m^2 \eta^{\la\mu} .
\eeal
The scalar~$Z$ implements the long-known fact~\cite{Bern:1993wt}
that two massless scalar states are produced
in the double copy of pure Yang-Mills theory
in addition to the two graviton states.
More precisely, the additional states correspond
to a dilaton scalar~$\phi$ and Kalb-Ramond two-form~$B_{\mu\nu}$.
In \app{app:fatgravity} we review the four-dimensional dualization
of the latter field to a real pseudoscalar~$a$,
which can then be combined with the real scalar~$\phi$
into a single complex axiodilaton scalar~$Z$. The resulting Lagrangian
for the massless sector of the double-copy theory is
\be
   {\cal L}_0 = -\frac{2}{\kappa^2} R
    + \frac{ \partial_\mu \bar{Z}\;\!\partial^\mu Z }
           { \big(1-\frac{\kappa^2\!}{4}\bar{Z}Z\big)^2 }
    = -\frac{2}{\kappa^2} R
    + g^{\mu\nu} \partial_\mu \bar{Z}\;\!\partial_\nu Z
      \bigg( 1 + \sum_{k=1}^\infty
                 \frac{\kappa^{2k}}{2^{2k}} (k+1) (\bar{Z}Z)^k \bigg) .
\label{MasslessLagrangian}
\ee

The fact that the vertices~\eqref{GravityVertexVectorZ} stay proportional to $m^2$
even off shell (instead of developing additional momentum dependence)
follows from inspecting the massless limit,
in which the $Z$ scalar should have a linearly realized U(1) symmetry (which is part of a larger U($N_f+1$) symmetry of the Luciani model~\cite{Luciani:1977hp}). Hence any U(1)-violating axiodilaton coupling must vanish in the massless limit. 

Now let us inspect the little-group-antisymmetrized double copies
that produce the massive scalar~$\varphi$. We see that
it naturally couples both to the graviton
and to the axiodilaton~$Z$:\footnote{The following identities
are helpful for considering antisymmetric double copies:
\be
   \begin{aligned}
   \ket{p^{[a}}_{\;\!\!\alpha}\;\![p^{b]}|_{\dot{\beta}} &
    = -\frac{1}{2} \epsilon^{ab} p_{\alpha\dot{\beta}} \, , \\
   |p^{[a}]^{\dot{\alpha}}\:\!\bra{p^{b]}}^{\beta} &
    = \frac{1}{2} \epsilon^{ab} p^{\dot{\alpha}\beta} ,
   \end{aligned} \qquad \quad
   \begin{aligned}
   \ket{p^{[a}}_{\;\!\!\alpha}\;\!\bra{p^{b]}}^\beta &
    = \frac{m}{2} \epsilon^{ab} \delta_\alpha^\beta \, , \\
   |p^{[a}]^{\dot{\alpha}}\;\:\!\! [p^{b]}|_{\dot{\beta}} &
    = -\frac{m}{2} \epsilon^{ab} \delta^{\dot{\alpha}}_{\dot{\beta}} \, .
   \end{aligned}
\label{AntisymProperties}
\ee
}
\begin{align}
   \frac{i\kappa}{2} A(1^{[a_1}\:\!\!,2^{[b_1}\:\!\!,3^+)
      A(1^{a_2]}\:\!\!,2^{b_2]}\:\!\!,3^+) &
    =-\frac{i\kappa}{4} \epsilon^{a_1 a_2} \epsilon^{b_1 b_2} m^2 x_3^2
    = \frac{1}{2} \epsilon^{a_1 a_2} \epsilon^{b_1 b_2}
      {\cal M}(1_{\varphi^*},2_\varphi,3^+) \, , \nn \\*
   \frac{i\kappa}{2} A(1^{[a_1}\:\!\!,2^{[b_1}\:\!\!,3^-)
      A(1^{a_2]}\:\!\!,2^{b_2]}\:\!\!,3^-) &
    =-\frac{i\kappa}{4} \epsilon^{a_1 a_2} \epsilon^{b_1 b_2} m^2 x_3^{-2}
    = \frac{1}{2} \epsilon^{a_1 a_2} \epsilon^{b_1 b_2}
      {\cal M}(1_{\varphi^*},2_\varphi,3^-) \, , \nn \\
   \frac{i\kappa}{2} A(1^{[a_1}\:\!\!,2^{[b_1}\:\!\!,3^\pm)
      A(1^{a_2]}\:\!\!,2^{b_2]}\:\!\!,3^\mp) &
    = \frac{i\kappa}{4} m^2 \epsilon^{a_1 a_2} \epsilon^{b_1 b_2}
    = \frac{1}{2} \epsilon^{a_1 a_2} \epsilon^{b_1 b_2}
      {\cal M}(1_{\varphi^*},2_\varphi,3_Z) \, .
\label{qcdDC3ptAntisym}
\end{align}
Here the last amplitude equals $i\kappa m^2/2$
and is due to the three-scalar coupling
\beal
\label{GravityVertexScalarZ}
   {\cal L}_{\varphi \varphi Z} &
    = \frac{\kappa}{2} (Z + \bar{Z}) m^2 \varphi^* \varphi
   \qquad \Rightarrow \quad~ 
   \scalegraph{0.9}{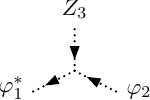}\!=\!\scalegraph{0.9}{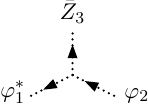}\!
    = \frac{i\kappa}{2} m^2 \,.
\eeal
Again, that these vertices stay proportional to $m^2$ off shell
follows from the fact that in the massless limit
the scalar $Z$ should have a conserved U(1) charge.

Finally, we should not forget about the combination
of little-group symmetrization and antisymmetrization
in the double copy:
\begin{align}
\label{qcdDC3ptMixed}
   \frac{i\kappa}{2} A(1^{(a_1}\:\!\!,2^{[b_1}\:\!\!,3^\pm)
      A(1^{a_2)}\:\!\!,2^{b_2]}\:\!\!,3^\pm) &
    = \frac{i\kappa}{2} A(1^{[a_1}\:\!\!,2^{(b_1}\:\!\!,3^\pm)
      A(1^{a_2]}\:\!\!,2^{b_2)}\:\!\!,3^\pm) = 0\, , \\*
   \frac{i\kappa}{2} A(1^{(a_1}\:\!\!,2^{[b_1}\:\!\!,3^+)
      A(1^{a_2)}\:\!\!,2^{b_2]}\:\!\!,3^-) &
    =-\frac{\kappa}{2\sqrt{2}}
      m (p_3\cdot\varepsilon_1^{a_1 a_2}) \epsilon^{b_1 b_2}
    = \frac{i\epsilon^{b_1 b_2}\!}{\sqrt{2}}
      {\cal M}(1_{V^*}^{a_1 a_2}\:\!\!,2_\varphi,3_Z) \,, \nn \\*
   \frac{i\kappa}{2} A(1^{(a_1}\:\!\!,2^{[b_1}\:\!\!,3^-)
      A(1^{a_2)}\:\!\!,2^{b_2]}\:\!\!,3^+) &
    =+\frac{\kappa}{2\sqrt{2}}
      m (p_3\cdot\varepsilon_1^{a_1 a_2}) \epsilon^{b_1 b_2}
    = \frac{i\epsilon^{b_1 b_2}\!}{\sqrt{2}}
      {\cal M}(1_{V^*}^{a_1 a_2}\:\!\!,2_\varphi,3_{\bar{Z}})\, , \nn \\*
   \frac{i\kappa}{2} A(1^{[a_1}\:\!\!,2^{(b_1}\:\!\!,3^+)
      A(1^{a_2]}\:\!\!,2^{b_2)}\:\!\!,3^-) &
    =-\frac{\kappa}{2\sqrt{2}}
      m (p_3\cdot\varepsilon_2^{b_1 b_2}) \epsilon^{a_1 a_2} 
    = \frac{i\epsilon^{a_1 a_2}\!}{\sqrt{2}}
      {\cal M}(1_{\varphi^*},2_V^{b_1 b_2}\:\!\!,3_Z)\, , \nn \\*
   \frac{i\kappa}{2} A(1^{[a_1}\:\!\!,2^{(b_1}\:\!\!,3^-)
      A(1^{a_2]}\:\!\!,2^{b_2)}\:\!\!,3^+) &
    =+\frac{\kappa}{2\sqrt{2}}
      m (p_3\cdot\varepsilon_2^{b_1 b_2}) \epsilon^{a_1 a_2} 
    = \frac{i\epsilon^{a_1 a_2}\!}{\sqrt{2}}
      {\cal M}(1_{\varphi^*},2_V^{b_1 b_2}\:\!\!,3_{\bar{Z}})\, . \nn
\end{align}
Here we see, as expected, that the gravitational interaction
does not mix massive scalars and vectors,
whereas the interaction with the axiodilaton does.
The latter non-trivial interaction is consistent with the following vertices
\beal
\label{ScalarVectorMixing}
   {\cal L}_{\varphi V Z} & = \frac{i \kappa}{4} m
      \big[
      (\varphi^* V_\mu + V^*_\mu \varphi)
      \partial^\mu (Z - \bar{Z})
    - (Z - \bar{Z})
      (\partial^\mu \varphi^*\;\!V_\mu + V^*_\mu \partial^\mu \varphi)
      \big] \\* & \Rightarrow~\left\{
   \begin{aligned}
   \scalegraph{0.9}{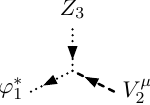}\!\! &= (-)\!\!\scalegraph{0.9}{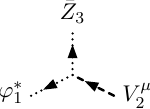}\!\!
    = -\frac{i\kappa}{4} m (p_1-p_3)^\mu , \\
   \scalegraph{0.9}{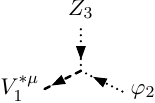}\!\! &= (-)\!\!\scalegraph{0.9}{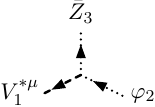}\!\!\!
    = -\frac{i\kappa}{4} m (p_2-p_3)^\mu ,
   \end{aligned} \right.
\eeal
where the relative signs are dictated by the hermiticity
of the resulting interactions with either complex or self-adjoint matter,
hence the $i$ prefactors in the double copies~\eqref{qcdDC3ptMixed}.

\section{Double copy for four-point vertices}
\label{sec:4pt}

In this section we transition to four-point amplitudes in gauge theory and gravity.
First we focus on Compton scattering,
by which we understand a process with two matter particles $1$ and $4$
interacting with two massless bosons $2$ and $3$.
The resulting four-point double copies are consistent
with the precise off-shell forms of the three-point vertices
given in the previous section.
Then we consider the pure-matter amplitudes,
for which we have to differentiate between the complex and self-conjugate matter,
as well as discuss the introduction of multiple flavors of such matter.

\subsection{Compton scattering in gauge theory}
\label{sec:compton}

First of all, we use the BCFW approach~\cite{Britto:2004ap,Britto:2005fq}
to derive the scattering amplitude for two massive particles and two gluons.
Assuming for concreteness the distinct-helicity case,
we shift the spinors of the gluons $2^+$ and~$3^-$ to
\be
   \ket{\hat{2}} = \ket{2} + z \ket{3} \, , \qquad \quad
   |\hat{3}] = |3] - z |2] \, .
\label{BCFWshift23}
\ee
We also assume a vanishing behavior at $z \to \infty$, which is guaranteed
for $s=0$ and $1/2$~\cite{Badger:2005zh,Badger:2005jv,Britto:2012qi}.
Instead of specializing to the latter case right away
(also considered in \rcite{Ochirov:2018uyq}),
let us leave the spin~$s$ of the massive matter unspecified for the moment,
which will make our discussion in \sec{sec:outro} more interesting.

There are two physical poles in $z$ that
come from the $s_{12}$- and $s_{13}$-channel propagators and localize at
\be
   z_{12} = -\frac{\bra{2}1|2]}{\bra{3}1|2]} \,, \qquad \quad
   z_{13} = \frac{\bra{3}1|3]}{\bra{3}1|2]} \, .
\label{BCFWshift23kinem}
\ee
These poles imply the following kinematic identities:
\begin{subequations} \begin{align}\!
   z_{12}: \quad &
   \hat{x}_2 
    = \frac{\bra{3}1|2]}{m\braket{2\;\!3}} \,, ~~
   \hat{x}_3^{-1}\!
    = \frac{\bra{3}1|2]}{m[3\;\!2]} \,, ~~~
   \bra{1^a}\hat{P}|4^b] = \frac{\!-m^2}{\bra{3}1|2]}
      \big( [1^a 2] \braket{3\;\!4^b} {+} \braket{1^a 3} [2\;\!4^b] \big) \,;\! \\*\!
   z_{13}: \quad &
   \hat{x}_2 
    = \frac{\bra{3}1|2]}{m\braket{2\;\!3}} \,, ~~
   \hat{x}_3^{-1}\!
    = \frac{\bra{3}1|2]}{m[3\;\!2]} \,, ~~~
   [1^a|\hat{P}\ket{4^b} = \frac{\!-m^2}{\bra{3}1|2]}
      \big( [1^a 2] \braket{3\;\!4^b} {+} \braket{1^a 3} [2\;\!4^b] \big) \, ,\!
\end{align} \label{BCFWshift23kinem2}%
\end{subequations}
where $\hat{P}$ is the cut momentum equal to
either $-(p_1+\hat{p}_2)$ or $-(p_1+\hat{p}_3)$, respectively.
Since on each of the poles the four-point amplitude
factorizes into two three-point amplitudes of the type~\eqref{qcdmatter3pt},
we compute the residues as
\begin{subequations} \begin{align}
\label{bcfwmmgg1}
   \scalegraph{0.9}{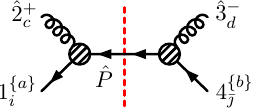} &
    = {\cal A}(1_i^{\{a\}}\:\!\!,\hat{2}_c^+\:\!\!,\hat{P}_{\bar k}^{\{e\}})
      \frac{(-1)^{\lfloor s \rfloor} i}{s_{12}-m^2}
      {\cal A}(-\hat{P}_{k\:\!\{e\}},\hat{3}_d^-\:\!\!,4_{\bar\jmath}^{\{b\}}) \\* &
    = \frac{ (-1)^{\lfloor s+1/2 \rfloor} ig^2 T^c_{i \bar k} T^d_{k\bar\jmath} }
           { (s_{12}-m^2) s_{23} }
      \bra{3}1|2]^{2-2s}
      \big( [1^a 2] \braket{3\;\!4^b}\!+\!\braket{1^a 3} [2\;\!4^b]
      \big)^{\odot 2s} , \nn \\
\label{bcfwmmgg2}
   \scalegraph{0.9}{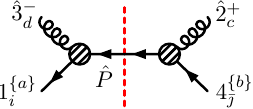} &
    = {\cal A}(1_i^{\{a\}}\:\!\!,\hat{3}_d^-\:\!\!,\hat{P}_{\bar k}^{\{e\}})
      \frac{(-1)^{\lfloor s \rfloor} i}{s_{13}-m^2}
      {\cal A}(-\hat{P}_{k\:\!\{e\}},\hat{2}_c^+\:\!\!,4_{\bar\jmath}^{\{b\}}) \\* &
    = \frac{ (-1)^{\lfloor s+1/2 \rfloor} ig^2 T^d_{i \bar k} T^c_{k\bar\jmath} }
           { (s_{13}-m^2) s_{23} }
      \bra{3}1|2]^{2-2s}
      \big( [1^a 2] \braket{3\;\!4^b}\!+\!\braket{1^a 3} [2\;\!4^b]
      \big)^{\odot 2s} . \nn
\end{align} \label{bcfwqqgg}%
\end{subequations}
Here we have also taken care of putting a spin-dependent sign in the propagator.
This is due to the fact that the completeness relations in the propagator numerators
has a sign-flip pattern $(-1)^{\lfloor s \rfloor}$ under
$\ket{{-}\hat{P}^e} = -\ket{\hat{P}^e}$, $|{-}\hat{P}^e] = |\hat{P}^e]$.
Indeed, for spins~$1/2$ and~$1$ we have
\be
   u_{p}^a \bar{v}_{-p\:\!a} = u_{p}^a \bar{u}_{p\:\!a} = \not{\!p}+m \, ,
   \qquad \quad
   \varepsilon_{p\:\!\mu}^{ab} \varepsilon_{-p\:\!\nu\:\!ab}
    = -\varepsilon_{p\:\!\mu}^{ab} \varepsilon_{p\:\!\nu\:\!ab}
    = -\bigg[{-\eta_{\mu\nu}} + \frac{p_\mu p_\nu}{m^2} \bigg] \,,
\ee
and so on for the higher-spin
external wavefunctions given in \eqns{poltensors}{polspinors}.

Therefore, the full color-dressed amplitude is
given by the sum of \eqns{bcfwmmgg1}{bcfwmmgg2}:\footnote{A formula
similar to \eqn{comptonQCD} was written in \rcite{Arkani-Hamed:2017jhn}
based on the residues of all three physical poles.}
\begin{align}
\label{comptonQCD}
 & {\cal A}(1^{\{a\}}_i\:\!\!,2^+_{c}\:\!\!,
            3^-_{d}\:\!\!,4^{\{b\}}_{\bar\jmath}) \\* &~
    = (-1)^{\lfloor s+1/2 \rfloor} ig^2\;\!\!
      \bigg[ \frac{T_{i \bar k}^c T_{k \bar\jmath}^d}{(s_{12}-m^2)s_{23}}
           + \frac{T_{i \bar k}^d T_{k \bar\jmath}^c}{(s_{13}-m^2)s_{23}}
      \bigg] \bra{3}1|2]^{2-2s}
      \big( [1^a 2] \braket{3\;\!4^b} {+} \braket{1^a 3} [2\;\!4^b]
      \big)^{\odot 2s} . \nn
\end{align}
The same method can be used to derive the identical-helicity amplitudes,
which are even simpler.
We find
\be
\label{comptonQCDallplus}
   {\cal A}(1^{\{a\}}_{i}\:\!\!,2^+_{c}\:\!\!,
            3^+_{d}\:\!\!,4^{\{b\}}_{\bar\jmath}) = (-1)^{\lfloor s \rfloor} ig^2
      \bigg[ \frac{T_{i \bar k}^c T_{k \bar\jmath}^d}{s_{12}-m^2}
           + \frac{T_{i \bar k}^d T_{k \bar\jmath}^c}{s_{13}-m^2}
      \bigg] \frac{ \braket{1^a 4^b}^{\odot 2s} [2\;\!3] }
                  { m^{2s-2} \braket{2\;\!3} } \,,
\ee
and ${\cal A}(1^{\{a\}}_i\:\!\!,2^-_c\:\!\!,3^-_d\:\!\!,4^{\{b\}}_{\bar\jmath})$
is the same up to the bracket swap $\braket{\:\!\dots} \leftrightarrow [\,\dots]$.

Now let us apply the standard adjoint-representation
color-ordering~\cite{Berends:1987cv,Mangano:1987xk,Mangano:1988kk}
to either of the above Compton amplitudes.
For instance, the all-plus amplitude~\eqref{comptonQCDallplus}
decomposes into three ordered amplitudes
\begin{align}
   A(1,2,3,4) &
    = \frac{(-1)^{\lfloor s \rfloor+1} i \braket{1^a 4^b}^{\odot 2s} [2\;\!3]^2}
           {m^{2s-2} (s_{12}\!-\!m^2) s_{23}}\, , \qquad \quad
   A(1,3,2,4)
    = \frac{(-1)^{\lfloor s \rfloor+1} i \braket{1^a 4^b}^{\odot 2s} [2\;\!3]^2}
           {m^{2s-2} (s_{13}\!-\!m^2) s_{23}} \,, \nn \\*
   A(1,3,4,2) &
    = \frac{(-1)^{\lfloor s \rfloor+1} i \braket{1^a 4^b}^{\odot 2s} [2\;\!3]^2}
           {m^{2s-2} (s_{12}\!-\!m^2)(s_{13}-m^2)}\, .
\label{comptonQCDpartials}%
\end{align}
We are now in a position to observe that, defined in this way for arbitrary spin,
the color-ordered amplitudes satisfy both the Kleiss-Kuijf (KK) relation~\cite{Kleiss:1988ne}
\be
   A(1,2,3,4) + A(1,3,2,4) + A(1,3,4,2) = 0
\ee
and the BCJ relation~\cite{Johansson:2015oia}
\be
   (s_{12}-m^2) A(1,2,3,4) = (s_{13}-m^2) A(1,3,2,4) \,.
\label{comptonBCJ}
\ee

\subsection{Compton scattering in gravity}
\label{sec:comptongravity}

Recall that the BCJ relations are a manifestation of the color-kinematics duality~\cite{Bern:2008qj,Bern:2010ue,Johansson:2014zca,Johansson:2015oia}, and the duality implies the BCJ double copy~\cite{Bern:2008qj}. The BCJ double copy is also known to be equivalent~\cite{Bern:2010yg}
to the tree-level KLT formulae~\cite{Kawai:1985xq} that are valid for external adjoint particles, or at most two non-adjoint ones~\cite{Johansson:2015oia}.
Therefore, the fact that the color-ordered amplitudes defined above
obey the BCJ relation~\eqref{comptonBCJ} means that
the gravitational Compton amplitudes are given
by the following gauge-invariant formula
\be
\label{comptonDC4pt}\!\!
   {\cal M}(1_s^{\{a\}}\:\!\!,2,3,4_s^{\{b\}})
    = (-1)^{\lfloor s \rfloor - \lfloor s_1 \rfloor - \lfloor s_2 \rfloor+1}
      i\Big(\frac{\kappa}{2}\Big)^{\!2}\!s_{23}\;\!
      A(1_{s_1}^{\{a\}}\:\!\!,2,3,4_{s_1}^{\{b\}})\odot
      A(1_{s_2}^{\{a\}}\:\!\!,3,2,4_{s_2}^{\{b\}}) \,,
\ee
where again the symmetrized tensor products builds up
gravitating matter of spin $s=s_1+s_2$
from two copies of QCD matter of spin $s_1$ and $s_2$.
We have included the same sign prefactor
that appears in the three-point double copy~\eqref{gravitymatter3ptKLT}.
This prescription implies
\bse
\begin{align}
\label{comptonGR1}
   {\cal M}(1_s^{\{a\}}\:\!\!,2^+\:\!\!,3^-\:\!\!,4_s^{\{b\}}) &
    = \Big(\frac{\kappa}{2}\Big)^{\!2}\!
      \frac{(-1)^{\lfloor s+1/2 \rfloor} i \bra{3}1|2]^{4-2s}}
           {(s_{12}\!-\!m^2)(s_{13}-m^2)s_{23}} 
      \big( [1^a 2] \braket{3\;\!4^b} + \braket{1^a 3} [2\;\!4^b]
      \big)^{\odot 2s} , \\
\label{comptonGR2}
   {\cal M}(1_s^{\{a\}}\:\!\!,2^+\:\!\!,3^+\:\!\!,4_s^{\{b\}}) &
    = \Big(\frac{\kappa}{2}\Big)^{\!2}\!
      \frac{(-1)^{\lfloor s \rfloor} i \braket{1^a 4^b}^{\odot 2s} [2\;\!3]^4}
           {m^{2s-4} (s_{12}-m^2) (s_{13}-m^2) s_{23}} \,.
\end{align} \label{comptonGR}%
\ese
The first of these expressions was earlier obtained in \rcite{Arkani-Hamed:2017jhn}
from matching to the residues of the three physical poles,
while to our knowledge the second has not previously appeared in the literature.
\Rcites{Arkani-Hamed:2017jhn,Chung:2018kqs} go into some detail
discussing the implications of the unphysical pole~$\bra{3}1|2]$
that appears in \eqn{comptonGR1} for $s>2$.
There is still a lot to be learned about massive higher-spin fields~\cite{Singh:1974qz,Singh:1974rc,Zinoviev:2002xn,Ondo:2016cdv,Buchbinder:2019dof}
from the on-shell perspective.
We postpone the discussion of the unphysical-pole issue at higher spins
to \sec{sec:outro}.
Instead, let us specialize to the case of $s=1$ constructed from two spins~$1/2$.
We find that for all helicity configurations
the little-group-symmetrized formulae~\eqref{comptonGR} agree
with the gravitational Compton amplitude obtained from the Feynman diagrams:
\begin{align}
\label{comptonGRvector}
   {\cal M}(1_{V^*}^{a_1 a_2},2,3,4_V^{b_1 b_2}) &
    = i\Big(\frac{\kappa}{2}\Big)^{\!2}\!s_{23}\;\!
      A(1^{(a_1}\:\!\!,2,3,4^{(b_1}) A(1^{a_2)}\:\!\!,3,2,4^{b_2)}) \\ &
    = \scalegraph{0.9}{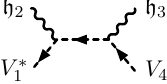} + \scalegraph{0.9}{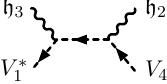}
    +\!\scalegraph{0.9}{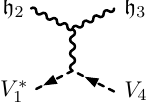}\!+ \scalegraph{0.9}{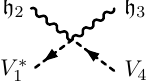} . \nn
\end{align}
Here we have used the four-point vertex implied by
the minimal coupling of the massive vector to gravity,
which is given explicitly in \eqn{GravityVertexVector2}
of \app{app:gravityvertices}.
The corresponding scalar amplitude is obtained
from ${\rm SU}(2)$ antisymmetrization
\be
\label{comptonGRscalar}
   {\cal M}(1_{\varphi^*},2,3,4_\varphi)
    = \frac{1}{2} \epsilon_{a_1 a_2} \epsilon_{b_1 b_2}
      \bigg[ {-i}\Big(\frac{\kappa}{2}\Big)^{\!2}\!s_{23}\;\!
             A(1^{a_1}\:\!\!,2,3,4^{b_1}) A(1^{a_2}\:\!\!,3,2,4^{b_2}) \bigg] ,
\ee
which is a four-point analogue of \eqn{qcdDC3ptAntisym}.
The resulting amplitude matches the Feynman-diagram calculation
involving the four-point minimal-coupling vertex~\eqref{GravityVertexScalar2}.

Choosing the helicities of one of the gluons 
in the double copies~\eqref{comptonGRvector} and~\eqref{comptonGRscalar}
to be anti-aligned results in the amplitudes
${\cal M}(1_{V^*},2_\hh,3_Z,4_V)$ and ${\cal M}(1_{\varphi^*},2_\hh,3_Z,4_\varphi)$,
which are consistent with the off-shell form of the three-point
vertices~\eqref{GravityVertexVectorZ} and~\eqref{GravityVertexScalarZ},
as well as their minimal four-point extensions
\begin{align}
\label{GravityVertexVectorGZ}
   \scalegraph{0.9}{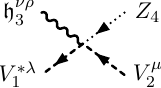} & = \scalegraph{0.9}{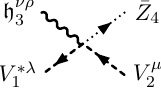}\!
    = \frac{i\kappa^2\!}{2} m^2 \eta^{\la(\nu} \eta^{\rho)\mu} , \\
\label{GravityVertexScalarGZ}
   \scalegraph{0.9}{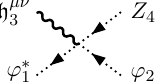} & =\:\!\scalegraph{0.9}{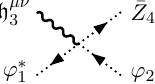}
    = -\frac{i\kappa^2}{4} m^2 \eta^{\mu\nu} ,
\end{align}
which come exclusively
from the nonlinearities in the $\hh^{\mu\nu}$-graviton expansion.

One should not forget that
the symmetric and antisymmetric matter double copies can be combined,
which for the three-point amplitudes we did in \eqn{qcdDC3ptMixed}.
Its generalization to four points is straightforward:
\bse
\begin{align}
   {\cal M}(1_{V^*}^{a_1 a_2}\:\!\!,2,3,4_\varphi) &
    = \frac{i\epsilon_{b_1 b_2}\!}{\sqrt{2}}
      \bigg[ {-i}\Big(\frac{\kappa}{2}\Big)^{\!2}\!s_{23}\;\!
             A(1^{(a_1}\:\!\!,2,3,4^{b_1}) A(1^{a_2)}\:\!\!,3,2,4^{b_2}) \bigg] , \\
   {\cal M}(1_{\varphi^*},2,3,4_V^{b_1 b_2}) &
    = \frac{i\epsilon_{a_1 a_2}\!}{\sqrt{2}}
      \bigg[ {-i}\Big(\frac{\kappa}{2}\Big)^{\!2}\!s_{23}\;\!
             A(1^{a_1}\:\!\!,2,3,4^{(b_1}) A(1^{a_2}\:\!\!,3,2,4^{b_2)}) \bigg] .
\end{align} \label{qcdDC4ptMixed}%
\ese
If the gluon helicities are aligned to produce gravitons,
these double copies continue to vanish,
which reinforces our claim that the vector~$V$ and the scalar~$\varphi$
are only mixed by their interaction with the axiodilaton~$Z$.
Indeed, switching one of the gravitons to either~$Z$ or~$\bar{Z}$
allows us to probe the following Feynman vertices,
\bse
\begin{align}
   \scalegraph{0.9}{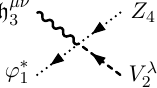}\!&= (-)\scalegraph{0.9}{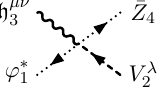}\!
    = \frac{i\kappa^2\!}{4} m \eta^{\la(\mu} [p_1-p_4]^{\nu)} \, , \\
   \scalegraph{0.9}{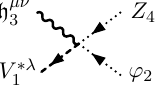}\!&= (-)\scalegraph{0.9}{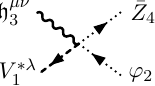}\:\!\!
    = \frac{i\kappa^2\!}{4} m \eta^{\la(\mu} [p_2-p_4]^{\nu)} \, ,
\end{align} \label{ScalarVectorMixing4}%
\ese
which are derived from the covariantization
of the scalar-vector mixing term that we proposed in \eqn{ScalarVectorMixing}.
Taking both massless particles to be scalars,
we see that the amplitudes
${\cal M}(1_{\varphi^*},2_Z,3_Z,4_V)$,
${\cal M}(1_{\varphi^*},2_{\bar{Z}},3_{\bar{Z}},4_V)$,
${\cal M}(1_{V^*},2_Z,3_Z,4_\varphi)$ and
${\cal M}(1_{V^*},2_{\bar{Z}},3_{\bar{Z}},4_\varphi)$
match corresponding Feynman-rules calculations automatically,
whereas the amplitudes
${\cal M}(1_{\varphi^*},2_Z,3_{\bar{Z}},4_V)$ and
${\cal M}(1_{V^*},2_Z,3_{\bar{Z}},4_\varphi)$
require an additional four-point vertex
\be
   \scalegraph{0.9}{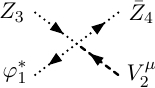} = \scalegraph{0.9}{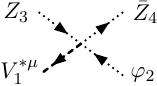}
    = -\frac{i\kappa^2\!}{4} m (p_3-p_4)^\mu \, .
\ee

Going back to the double copies~\eqref{comptonGRvector} and~\eqref{comptonGRscalar}
and also switching both massless legs to scalars,
we find that
${\cal M}(1_{V^*},2_Z,3_Z,4_V)$ and
${\cal M}(1_{V^*},2_{\bar{Z}},3_{\bar{Z}},4_V)$
each correspond to two trivalent Feynman diagrams, while 
${\cal M}(1_{V^*},2_Z,3_{\bar{Z}},4_V)$ calls for an additional quartic vertex
\be
   \scalegraph{0.9}{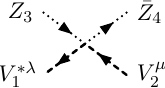}\!=
      \frac{i\kappa^2\!}{2} m^2 \eta^{\la\mu} \,.
\ee
Finally, the double copies for
${\cal M}(1_{\varphi^*},2_Z,3_Z,4_\varphi)$,
${\cal M}(1_{\varphi^*},2_{\bar{Z}},3_{\bar{Z}},4_\varphi)$ and
${\cal M}(1_{\varphi^*},2_Z,3_{\bar{Z}},4_\varphi)$
give us a handle on a new set of purely scalar vertices
\bse
\begin{align} \label{eqn4.21a}
   \scalegraph{0.9}{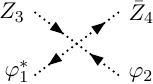}\!& =
     -\frac{i\kappa^2\!}{8} \big[(p_1\cdot p_2) + (p_3\cdot p_4) + 4m^2\big] \, , \\
   \scalegraph{0.9}{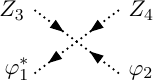} & = \scalegraph{0.9}{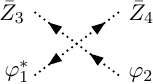}\!=
     -\frac{i\kappa^2\!}{8} m^2 \, .
\end{align}
\ese
Note that since the vertex~\eqref{eqn4.21a} contains elements
quadratic in both momenta and mass,
which are related due to the on-shell conditions of the four-point amplitude,
there is no unique way to write the off-shell vertex.
However, such ambiguities correspond to the freedom
to perform field redefinitions in the resulting Lagrangian,
which would affect the form of $(n \ge 5)$-point vertices
but not change the physics of the theory.
We may therefore proceed
by judiciously choosing an off-shell expression for this vertex. 
The explicit vertices given above correspond
to quartic terms in the Lagrangian~\eqref{PertLagrangian} given in \sec{sec:intro}.

\subsection{Four-matter vertices}
\label{sec:4matter}

In this section we turn to amplitudes with four matter particles.
In QCD, the distinct-flavor four-quark amplitude
${\cal A}(1^a_i,2^b_{\bar\jmath},3^c_k,4^d_{\bar l})$
consists of the sole gauge-invariant Feynman diagram
\be
   \scalegraph{0.9}{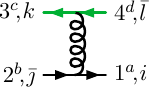}
    =-\frac{i T^e_{i \bar \jmath} T^e_{k \bar l}}{s_{12}}
      \big( \braket{1^a 4^d} [2^b 3^c]
          + [1^a 4^d] \braket{2^b 3^c}
          + \braket{1^a 3^c} [2^b 4^d]
          + [1^a 3^c] \braket{2^b 4^d}
      \big) \,,
\label{qqQQ}
\ee
where the masses and gauge-group representations of the two quark lines
may be different.
The identical-flavor amplitude is then a combination of two such diagrams
with the masses and representations taken equal
\be
   {\cal A}^{\rm id}(1^a_i,2^b_{\bar\jmath},3^c_k,4^d_{\bar l})
    = \scalegraph{0.9}{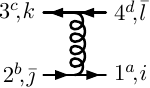} - \scalegraph{0.9}{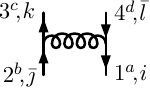}
    = {\cal A}(1^a_i,2^b_{\bar\jmath},3^c_k,4^d_{\bar l})
    - {\cal A}(1^a_i,4^d_{\bar l},3^c_k,2^b_{\bar\jmath}) \, ,
\label{qqqq}
\ee
where the relative sign is due to Fermi-Dirac statistics.
Now in the case of Majorana fermions,
the distinct-flavor amplitude stays the same as \eqn{qqQQ}.
That is, up to the switch to a real representation of the gauge group,
which we take to be the adjoint representation for simplicity.
Then the identical-flavor amplitude develops an additional pole
that was forbidden by charge conservation in the Dirac case:
\be
   {\cal A}^{\rm id}(1^a_i,2^b_j,3^c_k,4^d_l)
    = \frac{\tf^{aeb}\!\tf^{ced}\!}{s_{12}}
      n\!\left[\scalegraph{0.8}{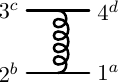}\right]
    - \frac{\tf^{aed}\!\tf^{ceb}\!}{s_{14}}
      n\!\left[\scalegraph{0.8}{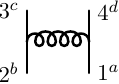}\right]
    + \frac{\tf^{aec}\!\tf^{deb}\!}{s_{13}}
      n\!\left[\scalegraph{0.8}{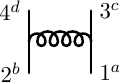}\right]\!,
\ee
\normalsize
where the kinematic dependence is still given by permutations of \eqn{qqQQ}.
Remarkably, a three-term identity for the kinematic numerators still holds
\be
   n\!\left[\scalegraph{0.9}{nmmMM}\right]
 + n\!\left[\scalegraph{0.9}{nmMMm}\right]
 + n\!\left[\scalegraph{0.9}{nmMmM}\right] = 0 \, ,
\label{BCJMajorana}
\ee
with the relative signs of the Jacobi identity modified by the fermionic statistics.
The BCJ relation
\be
   s_{12} A^{\rm id}(1^a\:\!\!,2^b\:\!\!,3^c\:\!\!,4^d) 
 - s_{13} A^{\rm id}(1^a\:\!\!,2^b\:\!\!,3^c\:\!\!,4^d) \xrightarrow[m\to0]{} 0 \,,
\ee
however, is only implied by the color-kinematics duality~\eqref{BCJMajorana}
in the massless case,
which is a well-known feature of supersymmetric Yang-Mills theories.

Now in this paper we exclusively rely on the color-kinematics duality of amplitudes
with distinctly flavored quarks~\cite{Johansson:2015oia,delaCruz:2015dpa}.
Other QCD amplitudes may be then obtained as linear combinations
of suitable relabelings thereof, as illustrated by \eqn{qqqq}.
In the same way, gravitational amplitudes with identically flavored matter
can be defined through linear combinations and relabelings of those with distinctly flavored matter.
So in the following we concentrate on how the latter are obtained
as double copies of distinct-flavor QCD amplitudes.

The fully little-group-symmetrized double copy
of the QCD amplitude~\eqref{qqQQ} gives
the distinct-flavor four-vector amplitude
that can be otherwise obtained
from the three-point Feynman vertices~\eqref{GravityVertexVector}
and~\eqref{GravityVertexVectorZ}
\begin{align}
   -i\Big(\frac{\kappa}{2}\Big)^{\!2}\!s_{34}
      A(1_{\bar{\Psi}}^{(a_1}\!,2_{\Psi}^{(b_1}\!,
        3_{\bar{\Psi}'}^{(c_1}\!,4_{\Psi'}^{(d_1}) &
      A(1_{\bar{\Psi}}^{a_2)}\!,2_{\Psi}^{b_2)}\!,
        4_{\Psi'}^{d_2)}\!,3_{\bar{\Psi}'}^{c_2)})
    = i\Big(\frac{\kappa}{2}\Big)^{\!2} \frac{1}{s_{12}}
      n\!\left[\scalegraph{0.8}{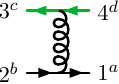}\right]^{\odot 2} \\*
    = {\cal M}(1_{V^*}^{a_1 a_2}\!,2_{V}^{b_1 b_2}\!,
               3_{V^{\prime *}}^{c_1 c_2}\!,4_{V'}^{d_1 d_2}) &
    = \scalegraph{0.9}{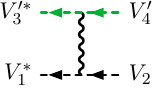} + \scalegraph{0.9}{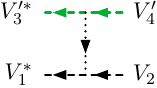}
    + \scalegraph{0.9}{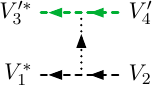} \, , \nn
\end{align}
confirming that there is no four-vector vertex
in the resulting gravitational theory.
On the other hand, the fully little-group-antisymmetrized double copy
results in such an amplitude for two massive scalar pairs\footnote{We use $M$
as a shorthand for gravitational amplitudes
without the trivial $i(\kappa/2)^{n-2}$ prefactor.
}
\be
   {-s_{34}}
      A(1_{\bar{\Psi}}^{[a_1}\!,2_{\Psi}^{[b_1}\!,
        3_{\bar{\Psi}'}^{[c_1}\!,4_{\Psi'}^{[d_1})
      A(1_{\bar{\Psi}}^{a_2]}\!,2_{\Psi}^{b_2]}\!,
        4_{\Psi'}^{d_2]}\!,3_{\bar{\Psi}'}^{c_2]})
    = \frac{1}{4} \epsilon^{a_1 a_2}  \epsilon^{b_1 b_2}
      \epsilon^{c_1 c_2} \epsilon^{d_1 d_2}
      M(1_{\varphi^*},2_{\varphi},3_{\varphi^{\prime *}},4_{\varphi'}) \, ,
\ee
which requires an additional four-scalar vertex
\be
   \scalegraph{0.9}{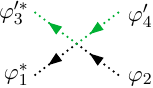}\!=-\frac{i\kappa^2\!}{8}
      \big[ (p_1 \cdot p_2) + (p_3 \cdot p_4) + 3m^2 +3m^{\prime 2} \big] \,.
\ee
Similarly to~\eqn{eqn4.21a}, here we have judiciously chosen an off-shell expression for the vertex that matches the on-shell contribution. 

Antisymmetrization of one or three pairs of little-group indices
yields vanishing amplitudes with one or three scalars,
in which only the axiodilaton may be exchanged:
\begin{align}\!
   {-s_{34}}
      A(1_{\bar{\Psi}}^{[a_1}\!,2_{\Psi}^{(b_1}\!,
        3_{\bar{\Psi}'}^{(c_1}\!,4_{\Psi'}^{(d_1})
      A(1_{\bar{\Psi}}^{a_2]}\!,2_{\Psi}^{b_2)}\!,
        4_{\Psi'}^{d_2)}\!,3_{\bar{\Psi}'}^{c_2)}) &
    = \frac{i\epsilon^{a_1 a_2}}{\sqrt{2}}
      M(1_{\varphi^*},2_{V}^{b_1 b_2}\!,
        3_{V^{\prime *}}^{c_1 c_2}\!,4_{V'}^{d_1 d_2}) \nn \\* &
    = \scalegraph{0.9}{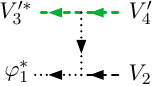} + \scalegraph{0.9}{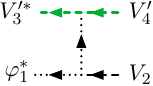} = 0 \,,  \nn\\*\!
   {-s_{34}}
      A(1_{\bar{\Psi}}^{(a_1}\!,2_{\Psi}^{[b_1}\!,
        3_{\bar{\Psi}'}^{(c_1}\!,4_{\Psi'}^{(d_1})
      A(1_{\bar{\Psi}}^{a_2)}\!,2_{\Psi}^{b_2]}\!,
        4_{\Psi'}^{d_2)}\!,3_{\bar{\Psi}'}^{c_2)}) &
    = \frac{i\epsilon^{b_1 b_2}}{\sqrt{2}}
      M(1_{V^*}^{a_1 a_2}\!,2_{\varphi},
        3_{V^{\prime *}}^{c_1 c_2}\!,4_{V'}^{d_1 d_2}) = 0 \, ,\;\\* 
   {-s_{34}}
      A(1_{\bar{\Psi}}^{[a_1}\!,2_{\Psi}^{[b_1}\!,
        3_{\bar{\Psi}'}^{[c_1}\!,4_{\Psi'}^{(d_1})
      A(1_{\bar{\Psi}}^{a_2]}\!,2_{\Psi}^{b_2]}\!,
        4_{\Psi'}^{d_2)}\!,3_{\bar{\Psi}'}^{c_2]}) &
    = \frac{i\epsilon^{a_1 a_2}  \epsilon^{b_1 b_2} \epsilon^{c_1 c_2}}{2\sqrt{2}}
      M(1_{\varphi^*},2_{\varphi},3_{\varphi^{\prime *}},
        4_{V'}^{d_1 d_2}) = 0 \, , \nn \\*
   {-s_{34}}
      A(1_{\bar{\Psi}}^{[a_1}\!,2_{\Psi}^{[b_1}\!,
        3_{\bar{\Psi}'}^{(c_1}\!,4_{\Psi'}^{[d_1})
      A(1_{\bar{\Psi}}^{a_2]}\!,2_{\Psi}^{b_2]}\!,
        4_{\Psi'}^{d_2]}\!,3_{\bar{\Psi}'}^{c_2)}) &
    = \frac{i\epsilon^{a_1 a_2}  \epsilon^{b_1 b_2} \epsilon^{d_1 d_2}}{2\sqrt{2}}
      M(1_{\varphi^*},2_{\varphi},
        3_{V^{\prime *}}^{c_1 c_2},4_{\varphi'}) = 0 \, . \nn
\end{align}
Note that in these amplitudes the two diagrams cancel
due to the sign difference
in the scalar-vector mixing vertices~\eqref{ScalarVectorMixing}.
Therefore, we see no need for vertices with more than two vector fields,
in accord with our expectations explained in \sec{sec:intro}.

Finally, symmetrizing two pairs of ${\rm SU}(2)$ indices
produces the following two-scalar two-vector gravitational amplitudes:
\beal\!\!\!
   {-s_{34}}
      A(1_{\bar{\Psi}}^{[a_1}\!,2_{\Psi}^{[b_1}\!,
        3_{\bar{\Psi}'}^{(c_1}\!,4_{\Psi'}^{(d_1})
      A(1_{\bar{\Psi}}^{a_2]}\!,2_{\Psi}^{b_2]}\!,
        4_{\Psi'}^{d_2)}\!,3_{\bar{\Psi}'}^{c_2)}) &
    =-\frac{1}{2} \epsilon^{a_1 a_2} \epsilon^{b_1 b_2}
      M(1_{\varphi^*},2_{\varphi},
        3_{V^{\prime *}}^{c_1 c_2}\!,4_{V'}^{d_1 d_2}) \, ,\!\! \\*\!\!\!
   {-s_{34}}
      A(1_{\bar{\Psi}}^{[a_1}\!,2_{\Psi}^{(b_1}\!,
        3_{\bar{\Psi}'}^{[c_1}\!,4_{\Psi'}^{(d_1})
      A(1_{\bar{\Psi}}^{a_2]}\!,2_{\Psi}^{b_2)}\!,
        4_{\Psi'}^{d_2)}\!,3_{\bar{\Psi}'}^{c_2]}) &
    =-\frac{1}{2} \epsilon^{a_1 a_2} \epsilon^{c_1 c_2}
      M(1_{\varphi^*},2_{V}^{b_1 b_2}\!,
        3_{\varphi^{\prime *}},4_{V'}^{d_1 d_2}) ,\!\!\!\\*\!\!\!
   {-s_{34}}
      A(1_{\bar{\Psi}}^{(a_1}\!,2_{\Psi}^{[b_1}\!,
        3_{\bar{\Psi}'}^{(c_1}\!,4_{\Psi'}^{[d_1})
      A(1_{\bar{\Psi}}^{a_2)}\!,2_{\Psi}^{b_2]}\!,
        4_{\Psi'}^{d_2]}\!,3_{\bar{\Psi}'}^{c_2)}) &
    =-\frac{1}{2} \epsilon^{b_1 b_2} \epsilon^{d_1 d_2}
      M(1_{V^*}^{a_1 a_2}\!,2_{\varphi},
        3_{V^{\prime *}}^{c_1 c_2}\!,4_{\varphi'}) \,, \\*\!\!\!
   {-s_{34}}
      A(1_{\bar{\Psi}}^{[a_1}\!,2_{\Psi}^{(b_1}\!,
        3_{\bar{\Psi}'}^{(c_1}\!,4_{\Psi'}^{[d_1})
      A(1_{\bar{\Psi}}^{a_2]}\!,2_{\Psi}^{b_2)}\!,
        4_{\Psi'}^{d_2]}\!,3_{\bar{\Psi}'}^{c_2)}) &
    =-\frac{1}{2} \epsilon^{a_1 a_2} \epsilon^{d_1 d_2}
      M(1_{\varphi^*},2_{V}^{b_1 b_2}\!,
        3_{V^{\prime *}}^{c_1 c_2}\!,4_{\varphi'}) \,,\!
\eeal
where only the first one allows graviton exchange
in addition to the axiodilaton exchange.
To coincide with the double copies above, these amplitudes require
supplementary mass-induced vertices:
\be\!
   \scalegraph{0.8}{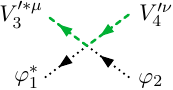}\!=
      \frac{i\kappa^2\!}{4} m^{\prime 2} \eta^{\mu\nu} , \quad
   \scalegraph{0.8}{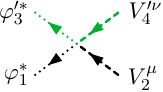}\!=\scalegraph{0.8}{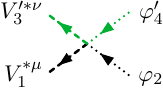}\!=\!\!
   \scalegraph{0.8}{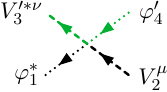}\!= \frac{i\kappa^2\!}{2} m m' \eta^{\mu\nu} .
\ee
This completes the list of vertices that could be determined from
the three- and four-point double copy
and enter the Lagrangian~\eqref{PertLagrangian} advertized in \sec{sec:intro}.
We leave exploring the higher-point vertices implied by the double copy
to future investigation.

\section{Gauge-invariant double copy}
\label{sec:bcj2klt}

In the previous sections we considered
three-point and distinct-flavor four-point double copies, for which
the transition between the BCJ and a KLT-like constructions was straightforward.
Inspecting the amplitude properties was sufficient
for identifying the necessary double-copy formulae,
and so the main focus there was on the explicit gravitational expressions.
In this section we switch focus to a more abstract approach for obtaining gauge-invariant double-copy formulae at any multiplicity.
This will reproduce the field-theory KLT relations~\cite{Kawai:1985xq},
but will also include a more general setup that
follows from the BCJ construction~\cite{Bern:2008qj}
using the color-kinematics duality and gauge-dependent numerators.
The extension~\cite{Johansson:2014zca,Johansson:2015oia} of the latter
to amplitudes involving massive and distinctly flavored matter
in arbitrary gauge-group representations
allows us to be very general in our discussion.

\subsection{Generalities of BCJ double copy}
\label{sec:bcj}

Consider a tree-level QCD amplitude~${\cal A}_{n,k}$ with $(n-2k)$ gluons
and $k$ distinctly flavored quark pairs.
The group-theoretic content of QCD involves only
structure constants~$\tf^{abc}$ and generators~$T^a_{i\bar\jmath}$,
which build up all the color factors $c_j$.
All Feynman diagrams can be grouped into
a distinct set of $(2n-5)!!/(2k-1)!!$ purely trivalent graphs, dictated by the color factors and flavor assignments~\cite{Johansson:2015oia}.
Now let us write the sum over these cubic graphs for three types of amplitudes:
for the $\phi^3$ theory of biadjoint and flavored bifundamental scalars (see
the Lagrangian~\eqref{BiScalarAction} below),
for QCD with flavored quarks and for gravity with flavored matter:\footnote{In
this section we remove the coupling constants
($y, g , \kappa/2 \to 1$). A factor of $i$ is omitted
from ${\cal A}^{\phi^3}_{n,k}$ and $M_{n,k}$,
along with the spin-dependent sign factor for $M_{n,k}$,
as they are irrelevant for the present discussion.
}
\be
   {\cal A}^{\phi^3}_{n,k} = \sum_{j} \frac{c_j \tilde{c}_j}{D_j} \, , \qquad \quad
   {\cal A}_{n,k} =  \sum_{j} \frac{c_j n_j}{D_j} \, , \qquad \quad
   M_{n,k} = \sum_{j} \frac{n_j \tilde{n}_j}{D_j} \, ,
\label{doublecopybcj}
\ee
where $D_j$'s are the propagator denominators corresponding to physical poles, and $n_j$ are the kinematic numerators that captures the non-trivial kinematic dependence in the QCD amplitudes. We also introduce tildes to allow repeated color factors and numerators to originate from different theories.
For the above connection between QCD and gravity to hold,
the kinematic numerators $n_j$ must
satisfy the same algebraic identities (Jacobi and commutation identities) as the color factors $c_j$,
\be
   c_i - c_j = c_k
   \qquad \Leftrightarrow \qquad
   n_i - n_j = n_k \,,
\label{bcjduality}
\ee
\ie they obey color-kinematics duality~\cite{Bern:2008qj,Johansson:2015oia}.
Hence the solution of the algebra of such identities
is the same for $c_j$ and $n_j$:
\beal
   c_j & = J_{j}^{~\bar\jmath} c_{\bar\jmath} , \qquad \quad\:\!
   j = 1, 2, \dots, (2n-5)!!/(2k-1)!! \quad \text{(all cubic diagrams)}\,, \\   
   n_j & = J_{j}^{~\bar\jmath} n_{\bar\jmath} , \qquad \quad
   \bar\jmath = 1, 2, \dots, (n-2)!/k! \quad\,\,~~~~~~~~~~~ \text{(basis of cubic diagrams)}\,,
\label{bcjduality2}
\eeal
where $J_{j}^{~\bar\jmath}$ is an integer-valued rectangular matrix obtained by repeatedly applying Jacobi/commutation relations until a basis of irreducible color factors have been obtained. The rank of the solution matrix $J_{j}^{~\bar\jmath}$ must equal $(n-2)!/k!$, since this is the size of the Melia basis of color-ordered amplitudes in QCD~\cite{Melia:2013bta,Melia:2013epa,Johansson:2015oia}. Without going into details, we denote the independent color-ordered amplitudes using the abstract notation
\be
   A_\alpha ,  \qquad 
   \alpha = 1, 2, \dots, (n-2)!/k!  \qquad \text{(KK/Melia basis)}\,,
\ee
where the indices~$\alpha$ can be understood as different cyclic orderings~\cite{Melia:2013bta,Melia:2013epa}.

The Melia basis is independent under a subset of the Kleiss-Kuijf relations~\cite{Kleiss:1988ne} that are compatible with the flavor assignments of the quarks.
In \rcite{Johansson:2015oia}, a new color decomposition was given for a color-dressed QCD amplitude in terms
of this basis of ordered amplitudes.\footnote{Alternative
amplitude bases and corresponding color decompositions can be found
in \rcite{Ochirov:2019mtf}.} The corresponding color coefficients $C^\alpha$ were given in terms of a explicit formula, and all the amplitudes in \eqn{doublecopybcj} can now be written as
\be
   {\cal A}_{n,k} = C^\alpha A_\alpha \qquad \Rightarrow \qquad
   {\cal A}^{\phi^3}_{n,k} = C^\alpha A_\alpha^{\phi^3} \, , \qquad \quad
   M_{n,k} = \tilde{K}^\alpha A_\alpha \,.
\label{doublecopy}
\ee
The color coefficients $C^\alpha$ are certain linear combinations of the cubic-graph color factors $c_j$, which we can write schematically as $C^\alpha= C^{\alpha \bar\jmath} c_{\bar\jmath}$. $C_{\alpha \bar\jmath}$ is an invertable square matrix that converts between $C^\alpha$'s and the basis of cubic-graph color factors $c_{\bar\jmath}$~\cite{Johansson:2015oia}.
The gravitational-amplitude kinematic coefficients~$\tilde K^\alpha$ are obtained from the color coefficients~$C^\alpha$
by swapping their constituent $c_j$'s by $\tilde n_j$'s, namely
\be
   C^\alpha = C^{\alpha \bar\jmath} c_{\bar\jmath} \, , \qquad \Rightarrow \qquad
   \tilde K^\alpha = C^{\alpha \bar\jmath} \tilde n_{\bar\jmath} \, ,
\label{colorfactormatrix}
\ee
Note that the above identity $M_{n,k} = \tilde{K}^\alpha A_\alpha$ is nothing but
our initial double-copy formula~\eqref{DCformula2015}~\cite{Johansson:2015oia},
rewritten in a more concise notation.

Let us now indirectly define a propagator matrix $P_{\alpha}^{~j}$
(see \eg \rcite{Vaman:2010ez}) for the color-ordered QCD amplitudes, by factoring out the numerators,  $A_\alpha = P_{\alpha}^{~j} n_j$.
Its rank must be $(n-2)!/k!$, otherwise the Melia basis would not be a basis,
so we can further reduce the numerators to the basis numerators~$n_{\bar\jmath}$:
\be
   A_\alpha = P_{\alpha}^{~j} n_j
    = P_{\alpha}^{~j}  J_{j}^{~\bar\jmath} n_{\bar\jmath} \qquad \Rightarrow \qquad
   A^{\phi^3}_\alpha = P_{\alpha}^{~j} c_j
    = P_{\alpha}^{~j}  J_{j}^{~\bar\jmath} c_{\bar\jmath} \, .
\label{qcdamplitudes}
\ee
Furthermore, inverting the color-factor matrix $C_{\alpha \bar\jmath}$
in \eqn{colorfactormatrix}, we obtain
\be
   A_\alpha = P_{\alpha}^{~j} J_{j}^{~\bar\jmath} C^{-1}_{\bar\jmath \beta}
              K^\beta .
\ee
In the purely gluonic case,
the resulting square matrix
$P_{\alpha}^{~j}  J_{j}^{~\bar\jmath} C^{-1}_{\bar\jmath \beta}$
turns out to be equal~\cite{Cachazo:2013iea}
to the sum over propagators of the graphs
occuring in both color-ordered amplitudes with subtle signs:\footnote{The
BCJ construction in this section
defines $A_{\alpha|\beta}$ with $\alpha$ and $\beta$
being permutations in a given Melia basis.
However, its amplitude interpretation~\eqref{propmatrixgraph}
immediately extends the definition to a pair of
arbitrary particle-label permutations $\alpha$ and $\beta$.
}
\be
   A_{\alpha|\beta} =
   P_{\alpha}^{~j}  J_{j}^{~\bar\jmath} C^{-1}_{\bar\jmath \beta}
    = (-1)^{n_\text{flip}(\alpha,\beta)}\!\!\!
      \sum_{\text{graph }j \in A_\alpha \cap \tilde{A}_\beta}
         \frac{1}{D_j} \, .
\label{propmatrixgraph}
\ee
It also coincides with the doubly color-ordered amplitude
for a theory of biadjoint and flavored bifundamental scalars.
To prove \eqn{propmatrixgraph}, it suffices to consider
the zeroth-copy version of the double copy~\eqref{doublecopy}
\be
   {\cal A}^{\phi^3}_{n,k} = C^\alpha A_\alpha^{\phi^3}
    = C^\alpha P_{\alpha}^{~j} J_{j}^{~\bar\jmath} C^{-1}_{\bar\jmath \beta}
      C^{\beta}
    = C^\alpha A_{\alpha|\beta} C^\beta ,
\ee
where the last expression is a double color decomposition.
An efficient way for computing this amplitude matrix is discussed below in \sec{sec:kltberendsgiele}.
The BCJ double copy in \eqn{doublecopy} can therefore be rewritten as~\cite{Johansson:2015oia}
\be
   M_{n,k} = K^\beta \tilde{A}_{\beta} \, , \qquad \quad
   A_\alpha = A_{\alpha|\beta} K^{\beta} \, , \qquad \quad
   {\rm rank}\;\!A_{\alpha|\beta} = (n-3)!(2k-2)/k! \, ,
\label{doublecopy2}
\ee
where we also allowed the second copy to be in a different theory
(with a matching flavor pattern),
and $\beta$ can in principle run over a different Melia-like basis.

\subsection{BCJ relations and KLT-like double copy}
\label{sec:klt}

The rank of the matrix $A_{\alpha|\beta}$
is in general smaller than the Melia basis size.
Assuming that color-kinematics duality~\eqref{bcjduality} holds,
we need to ensure that the linear system $A_{\alpha|\beta} K^{\beta} = A_\alpha$
is not inconsistent.
The Kronecker-Rouch\'e-Capelli theorem in linear algebra then implies
consistency conditions on the non-homogeneity of that linear equation
\be
   A_{\underline{\alpha}}
    = F_{\underline{\alpha}}^{~\bar{\alpha}} A_{\bar\alpha} , \qquad 
   \begin{aligned}
   \bar{\alpha} & = 1, 2, \dots, (n{-}3)!(2k{-}2)/k! \qquad\qquad~~\;\,
   \text{(BCJ basis)}\,, \\
   \underline{\alpha} & = (n{-}3)!(2k{-}2)/k!{+}1, \dots, (n{-}2)!/k!  \quad
   \text{(outside BCJ basis)}\,,
   \end{aligned}
\label{bcjrelations0}
\ee
which coincide with non-trivial BCJ relations.
Indeed, we can write a BCJ relation~\cite{Bern:2008qj,Johansson:2015oia}
schematically as
\be
   A_{\alpha} = F_{\alpha}^{~\bar{\alpha}} A_{\bar\alpha} \, .
\label{bcjrelations}
\ee
It expresses an arbitrary color-ordered amplitude $A_{\alpha}$ in the Melia basis
in terms of the amplitudes $A_{\bar\alpha}$ in the BCJ basis
with the kinematic coefficients $F_{\alpha}^{~\bar{\alpha}}$.
For the BCJ basis itself we have simply
$F_{\bar\alpha}^{~\bar\gamma}=\delta_{\bar\alpha}^{\bar\gamma}$.\footnote{Here
we assume that the BCJ basis is chosen as a subset of the Melia basis,
but in principle we can use Kleiss-Kuijf relations~\cite{Kleiss:1988ne}
to extend the BCJ relation matrix $F_{\alpha}^{~\bar{\beta}}$
to more general basis choices.
} 
The non-trivial part of the BCJ relations~\eqref{bcjrelations}
is of course given by \eqn{bcjrelations0}:
it resolves the non-BCJ-basis amplitudes in terms of the BCJ basis.
Plugging the consistency condition~\eqref{bcjrelations}
into our linear equation $A_{\alpha|\beta} K^{\beta} = A_\alpha$ we find
\be
   A_{\alpha|\beta} K^{\beta} = F_{\alpha}^{~\bar\gamma} A_{\bar\gamma}
    = F_{\alpha}^{~\bar\gamma} A_{\bar \gamma|\beta} K^{\beta} \, .
\ee
As this linear system has not yet been solved for $K^{\beta}$,
they can be regarded as arbitrary, and we conclude that
\be
   A_{\alpha|\beta} = F_{\alpha}^{~\bar\gamma} A_{\bar \gamma|\beta}
    = A_{\alpha|\bar \delta} F_{\beta}^{~\bar\delta} \, .
\label{bcjmatrixproperty}
\ee
In other words, the BCJ-relation matrix
$(F_{\underline{\alpha}}^{~\bar\gamma}\!-\delta_{\underline{\alpha}}^{\gamma})$
spans the kernel subspace of $A_{\alpha|\beta}$.

Now if we restrict the doubly color-ordered amplitude matrix~$A_{\alpha|\beta}$
to a square submatrix related to two sets of BCJ basis orderings,
we can define the momentum kernel as its inverse,
as was done already in \rcite{delaCruz:2016wbr}.
We have
\be
   S^{\bar\alpha \bar\beta} \equiv
       A^{-1\:\!\bar\beta|\bar\alpha}
    = (A^{-1\mathsf{T}})^{\bar\alpha|\bar\beta} , \qquad ~~~
   \bar{\alpha}, \bar{\beta} = 1, 2, \dots, (n-3)!(2k-2)/k! \quad
   \text{(BCJ basis)} \,.
\label{momentumkernel}
\ee
Then multiplication of \eqn{bcjmatrixproperty}
by the momentum kernel on the left and right gives the following formulae
for the BCJ-relation matrix~$F$:
\be
   F_{\alpha}^{~\bar\gamma}
    = A_{\alpha|\bar\beta} S^{\mathsf{T}\:\!\bar\beta \bar\gamma}\, , \qquad \quad
   F_{\beta}^{~\bar\delta}
    = S^{\mathsf{T}\:\!\bar\delta \bar\alpha} A_{\bar\alpha|\beta} \, .
\label{bcjmatrix2kltmatrix}
\ee

We can now go back to the double copy~\eqref{doublecopy2}
and solve the linear equation $A_{\bar\alpha|\beta} K^{\beta} = A_{\bar\alpha}$
for a subset of $K^{\bar\beta}$ corresponding to the second BCJ basis
\be
   K^{\bar\beta} = A^{-1\:\!\bar\beta|\bar\alpha}
      (A_{\bar\alpha} - A_{\bar\alpha|\underline{\delta}} K^{\underline{\delta}})
    = A_{\bar\alpha} S^{\bar\alpha \bar\beta}
    - F_{\underline{\delta}}^{~\bar\beta} K^{\underline{\delta}} \, .
\ee
Inserting it into the double-copy relation
$M_{n,k}=K^{\bar\beta} \tilde{A}_{\bar\beta}
    + K^{\underline{\beta}} \tilde{A}_{\underline{\beta}}$,
we see that the double copy takes a KLT-like form\footnote{There is
a good reason why the gauge-invariant double copy~\eqref{doublecopyklt}
looks like a symmetric bilinear form.
A basis change from one pair $A_{\bar\alpha}$, $\tilde{A}_{\bar\beta}$
of BCJ bases to another such pair $A_{\tilde\gamma}$, $\tilde{A}_{\tilde\delta}$
can be described by a generalized BCJ-relation matrix,
that allows for Kleiss-Kuijf relations as well.
The gravity amplitude is invariant with respect to such basis changes:
\be
   \left\{
   \begin{aligned}
   A_{\bar\alpha} & = F_{\bar\alpha}^{~\tilde\gamma} A_{\tilde\gamma}
    = A_{\bar\alpha|\tilde\zeta} A^{-1\:\!\tilde\zeta|\tilde\gamma}
      A_{\tilde\gamma} \\
   \tilde{A}_{\bar\beta} & = F_{\bar\beta}^{~\tilde\delta} \tilde{A}_{\tilde\delta}
    = \tilde{A}_{\tilde\delta} A^{\mathsf{T}\:\!\tilde\delta|\tilde\varepsilon}
      A_{\tilde\varepsilon|\bar\beta}
   \end{aligned}
   \right. \qquad \Rightarrow \qquad
   \begin{aligned}
   M_{n,k} & = A_{\bar\alpha} S^{\bar\alpha \bar\beta} \tilde{A}_{\bar\beta}
    = A_{\tilde\gamma}
      A^{-1\:\!\tilde\delta|\tilde\varepsilon} A_{\tilde\varepsilon|\bar\beta}
      A^{-1\:\!\bar\beta|\bar\alpha}
      A_{\bar\alpha|\tilde\zeta} A^{-1\:\!\tilde\zeta|\tilde\gamma}
      \tilde{A}_{\tilde\delta} \\ &
    = A_{\tilde\gamma} S^{\:\!\tilde\gamma \tilde\delta} \tilde{A}_{\tilde\delta} \,.
   \end{aligned}
\label{bcjbasischange}
\ee
Here we have used of a non-trivial property of the doubly color-ordered amplitude,
which follows from \eqns{bcjmatrixproperty}{bcjmatrix2kltmatrix},
\be
   A_{\alpha|\beta} = A_{\alpha|\bar\delta} A^{-1\:\!\bar\delta|\bar\gamma}
                      A_{\bar\gamma|\beta}
    = A_{\alpha|\bar\delta} S^{\mathsf{T}\:\!\bar\delta \bar\gamma}
                      A_{\bar\gamma|\beta} \, .
\label{kltampproperty}
\ee
} \cite{delaCruz:2016wbr}
\be
   M_{n,k} 
    = A_{\bar\alpha} S^{\bar\alpha \bar\beta} \tilde{A}_{\bar\beta} \, .
\label{doublecopyklt}
\ee

The double copy (\ref{doublecopyklt}) has the nice feature that it is manifestly gauge-invariant, but as a consequence it does not make manifest the simple poles and crossing symmetry of the gravitational amplitude. This is in contradistinction to the original BCJ double copy~\eqref{doublecopybcj}, where the poles and crossing symmetry are both manifest but gauge (diffeomorphism) invariance is not manifest. Interestingly, the double copy in~\eqn{DCformula2015} has both part of the gauge invariance manifest and the simple poles are also obvious from the gauge-theory amplitude, but crossing symmetry and gauge invariance of the kinematic factor is absent.

\subsection{Recursion for bi-colored scalar amplitudes}
\label{sec:kltberendsgiele}

The gauge-invariant form of the double copy~\eqref{doublecopyklt}
involves the matrix~$S^{\bar\alpha \bar\beta}$,
defined in \eqn{propmatrixgraph}
as the inverse of doubly color-ordered amplitudes in a theory of biadjoint and flavored bi-fundamental scalars.
Its Lagrangian, with symmetry group $G \times G'$, is explicitly
\beal
\label{BiScalarAction}
   {\cal L}_{\phi^3} &
    = \frac{1}{2} \partial_\mu \phi^{aA}\;\!\partial^\mu \phi^{aA}
    + \frac{y}{3!} \tf^{abc} \tf^{ABC} \phi^{aA} \phi^{bB} \phi^{cC} \\ & \quad
    + \sum_{\fla = 1}^{N_f} \Big(
      \partial_\mu \varphi_{\fla\:\!\bar\imath \bar I}^*\;\!
      \partial^\mu \varphi_{\fla\:\!i I}^{\phantom *}
    - m_\fla^2 \varphi_{\fla\:\!\bar\imath \bar I}^*
      \varphi_{\fla\:\!i I}^{\phantom *}
    + y \phi^{aA} \varphi_{\fla\:\!\bar\imath \bar I}^*
      T^a_{i \bar \jmath} T^A_{I \bar J}
      \varphi_{\fla\:\!j J}^{\phantom *} \Big) \,,
\eeal
where $\fla$ is a flavor index,
$(a, i, \bar\imath)$ are the (adjoint, fundamental, antifundamental) indices of the first group $G$,
and $(A, I, \bar I)$ are the corresponding indices for the second group $G'$.
This Lagrangian only produces trivalent Feynman diagrams
that encodes the same color and propagator structure as QCD.
An efficient way to compute amplitudes in this theory is
via the Berends-Giele recursion for the doubly color-ordered current:
\beal
   J_{\alpha|\beta} = \left\{
   \begin{aligned}
    & 0 \qquad \qquad \qquad \qquad \qquad \qquad \qquad \qquad~~\;\quad
      \text{ if } \alpha \setminus \beta \neq \emptyset , \\
    & \frac{\theta_\alpha}{s_\alpha}
         \sum_{\sigma\oplus\tau=\alpha} \sum_{\phi\oplus\psi=\beta}
         \Big( J_{\sigma|\phi} J_{\tau|\psi} - J_{\tau|\phi} J_{\sigma|\psi}
         \Big) \qquad
      \text{ otherwise, }
   \end{aligned} \right.\!
\label{doublecurrentrecursion}
\eeal
where we use $\oplus$ to denote
concatenation of two shorter particle-label permutations.
As indicated, the current vanishes unless $\alpha$ is a permutation of $\beta$.
The starting seed of the recursion is simply $J_{i|j} = \delta_{ij}$.
This recursion is different from
the formula written in \rcite{Mafra:2016ltu} (for the purely adjoint case)
by the $\theta$ factor, the purpose of which
is to eliminate illegal flavor assignments.\footnote{A consistent way
to implement the flavor $\theta$~function
is by introducing a symbolic flavor function~$f(\alpha)$ 
defined for a permutation $\alpha$ as a sum of flavor symbols of each particle,
 $ f(\alpha) = \sum_{i\in\alpha} f(i) $.
Assuming that, as in \rcite{Johansson:2015oia},
the first $k$ pairs of particle labels are assigned
to the distinctly flavored matter particles and the remaining $(n-2k)$
labels correspond to the massless interaction particles, we specify
\beal
 & f(2j-1) = -f_j \, , \qquad \quad
   f(2j) = f_j \, , \qquad \quad j=1,2,\dots,k \, , \\
 & f(i) = 0\, , \qquad \qquad \qquad \qquad \qquad \qquad \qquad~~
   i = (2k+1),(2k+2),\dots,n \, .
\eeal
Here $f_i \neq f_j$ are non-numeric symbols that stay uncanceled unless
two matter particles of the same flavor meet.
Then we can define
\be
   \theta_\alpha = \left\{
   \begin{aligned}
    & 1 \qquad \text{ if }f(\alpha)\text{ is a single term,}\\
    & 0 \qquad \text{ otherwise.}
   \end{aligned} \right.
\label{flavortheta}
\ee
For instance, $f(\o{1},\u{2},5,\u{4}) = -f_1+f_1+0+f_2=f_2$
is a single term, whereas $f(\o{1},5,\u{4}) =-f_1+f_2$ is not.
}

The doubly color-ordered amplitudes are obtained from currents
by removing the final denominator
and putting the corresponding external momentum on shell:
\be   
   A_{\alpha \oplus\{n\}|\beta \oplus\{n\}}
    = s_\alpha J_{\alpha|\beta}\,\big|_{s_\alpha = p_n^2 = 0} \, .
\label{doublecurrent2amplitude}
\ee
Therefore, the recursion relation given above provides an efficient way
for numerical evaluation of gauge-invariant double copies.

In \eqns{doublecurrentrecursion}{doublecurrent2amplitude} above
and in the rest of this paper,
we use the massless Mandelstam invariants
$s_\alpha=\big(\sum_{i\in\alpha} p_i\big)^2$ for simplicity.
The correct way to extend them to the massive case
is by switching to a mass-deformed notation
\be
   s_\alpha \to t_\alpha =\!\sum_{i<j\in\alpha\!\!}\!t_{ij}\, , \qquad \quad
   t_{ij} = \left\{
   \begin{aligned}
    & 2p_i \cdot p_j + 2m_i^2 \qquad
      \text{ if }i\text{ and }j\text{ of the same flavor,} \\
    & 2p_i \cdot p_j \qquad \qquad~\,\quad \text{ otherwise.}
   \end{aligned} \right.
\ee

\subsection{Four-matter double copies at arbitrary multiplicity}
\label{sec:KLTk2}

In this section we specialize to the case
of two distinctly flavored pairs of matter particles.
This is the simplest instance where the gauge-invariant double copy
non-trivially differs from the purely adjoint KLT double copy.
Similarly to \sec{sec:4matter},
we label the matter particles of the first flavor as $\o{1}$ and $\u{2}$
and those of the second flavor as $\o{3}$ and $\u{4}$.
The rest of the labels $5,\dots,n$ are gluons which double-copy
to gravitons or axiodilatons.
Then the Melia-type bases have a simple form of $A(\o{1},\u{2},\alpha)$,
with the relative ordering of $\o{3}$ and $\u{4}$ inside $\alpha$
fixed to either $\o{3}\leftarrow\u{4}$
or $\u{4}\rightarrow\o{3}$~\cite{Melia:2013bta,Melia:2013epa}.
A BCJ basis is then imposed by requiring that one of the previously unfixed quarks
be adjacent to the first pair, as in \eg $A(\o{1},\u{2},\o{3},\alpha)$,
where $\alpha$ is an arbitrary permutation
of the remaining particle labels~\cite{Johansson:2015oia}.

In \sec{sec:bcj2klt} we assumed the same BCJ basis
for both sides of the gauge-invariant double copy~\eqref{doublecopyklt}.
Such a choice, however, leads to non-localities in the momentum kernel.
This is actually true even in the flavorless case,
\ie for $A(1,2,3,\alpha) \times \tilde{A}(1,2,3,\beta)$.
In fact, even the purely adjoint KLT double copy
is realized non-locally for the BCJ basis combinations of the form
$A(1,2,\alpha,3,\beta) \times \tilde{A}(1,\gamma,2,3,\delta)$
with fixed sizes of $\alpha$, $\beta$, $\gamma$ and $\delta$,
unless at least one of them is set to zero.
This can be explained by the fact all the involved primitives
miss some physical poles, such as $s_{13}$, which have to occur in the gravity amplitude and therefore must be introduced by the momentum kernel.
A similar analysis implies that local combinations of flavored BCJ bases
require that the gluon-free slots be adjacent to the quark pair
fixed by the Melia basis,
but it cannot be simply
$A(\o{1},\u{2},\o{3},\alpha) \times A(\o{1},\u{2},\o{3},\beta)$,
as then neither side contains the $s_{2n}$-channel pole (where $n \geq 5$).

The simplest gauge-invariant double copy that we managed to find is
\beal
   M_{n,k=2} = \,\,\, - \!\!\!\!\!\!\!
      \sum_{\alpha,\beta \in S_{n-3}(\{4,\dots,n\})} \!\!\!\!
      A(\o{1},\u{2},\o{3},\alpha)
      \tilde{A}(\u{2},\o{1},\o{3},\beta)
      \prod_{i=1}^{n-3}
      \Big(\!\!\sum_{\substack{ j \in \{3,\alpha_1,\dots,\alpha_{i-1}\} \\
                            ~\;\cap \{3,\beta_{1},\dots,\beta_{b_i-1}\} }} \!\!\!\!
            s_{\alpha_i j}
      \Big) ,
\label{KLTk2flip}
\eeal
where $b_i$ is defined as the position of $\alpha_i$ within $\beta$, \ie
 $ \beta_{b_i} = \alpha_i $.
In fact, this basis combination corresponds
to a momentum kernel that mimics precisely the unflavored one~\cite{Brown:2018wss}.
For instance, it satisfies the same recursion relation
that was previously noticed in \rcite{Carrasco:2016ldy}:
\be
   S[\alpha|\beta] = - \prod_{i=1}^{n-3}
      \Big(\!\!\sum_{\substack{ j \in \{3,\alpha_1,\dots,\alpha_{i-1}\} \\
                  ~\;\cap \{3,\beta_{1},\dots,\beta_{b_i-1}\} }}\!\!\!\!
            s_{\alpha_i j}
      \Big) \quad \Rightarrow \quad
   S[\alpha,j|\beta,j,\gamma]
    = \Big( \sum_{i\in\{3,\beta_1,\dots,\beta_{n-3}\}}\!\!\!s_{\alpha_i j} \Big)
      S[\alpha|\beta,\gamma] \,.
\ee

Going to other gauge-invariant double-copy representations, however,
can be quite non-trivial,
as it involves a combination of flavored KK and BCJ relations
on one of the sides of the double copy.
For instance, another formula for two matter particle pairs reads
\bse \label{KLTk2}
\beal
   M_{n,k=2} = \!\!\!
      \sum_{\substack{\alpha \in S_{n-3}(\{4,5,\dots,n\}) \\
                      \beta \in S_{n-3}(\{3,5,\dots,n\})}} \!\!\!\!
      A(\o{1},\u{2},\o{3},\alpha)
      \tilde{A}(\o{1},\u{2},\beta,\u{4})\,
      (-1)^{n+|\gamma^*|}\,\theta_{34} & \\ \times
      \prod_{i=1}^{|\gamma^*|-1}\!s_{\gamma^*_i,\gamma^*_{i+1}}
      \prod_{i=1}^{|\gamma|}
      \Big(
            \sum_{\substack{ j \in \{3,\alpha_1,\dots,\alpha_{a_i-1}\}~~~\;\,\\
                              \cap \{\beta_{b_i+1},\dots,\beta_{n-3},4\},\\\,\,
                      \alpha_{a_i} = \beta_{b_i} = \gamma_i }}\!\!\!\!\!\!\!\!\!\!
            s_{\gamma_i j}
      \Big) & ,
\label{KLTk2main}
\eeal
where the gluon set $\gamma$ is defined as a complement
of the ordered subset $\gamma^*$,
which picks some of the gluons trapped inside bracket $\{3\dots4\}$
both in $\alpha$ and $\beta$:
\beal
   \gamma^* & = \{ \alpha_i : \{3,\alpha_1,\dots,\alpha_{i-1}\}
                         \cap \{\beta_{b_i+1},\dots,\beta_{n-3},4\} = \emptyset ,~
                         \beta_{b_i} = \alpha_i
                \} \setminus \{4\} \, , \\
   \gamma ~ & = \alpha \setminus (\gamma^* \oplus \{4\}) \,,
\eeal
and the prefactor $\theta_{34}$ is
\be
   \theta_{34} = \left\{
   \begin{aligned}
    & s_{34} \qquad \qquad \quad \text{ if } \gamma^* = \emptyset , \\
    & s_{3 \gamma^*_1} s_{4 \gamma^*_{|\gamma^*|}} \qquad \text{ otherwise.}
   \end{aligned} \right.
\label{KLTk2prefactor}
\ee
\ese

Another interesting basis choice for $k=2$ is obtained by fixing two different quark pairs on different sides of the double copy:
\bse \label{KLTk2compact}
\begin{align}
\label{KLTk2compactmain}
   M_{n,k=2} = & \!\!\!\!
      \sum_{\substack{\alpha \in S_{n-3}(\{4,5,\dots,n\}) \\
                      \beta \in S_{n-3}(\{2,5,\dots,n\})}} \!\!\!\!
      A(\o{1},\u{2},\o{3},\underbrace{\pi,\u{4},\sigma}_{\alpha})
      \tilde{A}(\o{3},\u{4},\o{1},\underbrace{\rho,\u{2},\tau}_{\beta})\,
      (-1)^{n+|\gamma|+\theta(|\gamma^*|>0)}~\theta_{31} \\ \times &
      \prod_{i=1}^{|\gamma^*|-1}\!s_{\gamma^*_i,\gamma^*_{i+1}}
      \prod_{i=1}^{|\gamma|}
      \Big(
            \sum_{\substack{ j \in \{\pi_1,\dots,\pi_{a_i-1}\}\\\;\:
                              \cap \{\rho_1,\dots,\rho_{b_i-1}\},\\~~~\;
                             \pi_{a_i} = \rho_{b_i} = \gamma_i }} \!\!\!\!\!
            s_{\gamma_i j}
      \Big)
      \prod_{i=1}^{|\sigma|}
      \Big(
            \sum_{\substack{ j \in \{\sigma_{i+1},\dots,\sigma_{|\sigma|},1\}\\\;
                              \cap \{1,\rho_1,\dots,\rho_{b_i-1}\}\\\;\!\!
                             \rho_{b_i} = \sigma_i }} \!\!\!\!\!
            s_{\sigma_i j}
      \Big)
      \prod_{i=1}^{|\tau|}
      \Big(
            \sum_{\substack{ j \in \{\tau_{i+1},\dots,\tau_{|\tau|},3\}\\\;
                              \cap \{3,\pi_1,\dots,\pi_{a_i-1}\}\\\:\!\!\!
                             \pi_{a_i} = \tau_i }} \!\!\!\!\!
            s_{\tau_i j}
      \Big) , \nn
\end{align}
where auxiliary gluon subsets $\gamma^*$ (ordered) and $\gamma$ (unordered)
are defined as
\beal
   \gamma^* & = \{ \pi_i : \{\pi_1,\dots,\pi_{i-1}\}
                         \cap \{\rho_1,\dots,\rho_{b_i-1}\} = \emptyset ,~
                         \rho_{b_i} = \pi_i
                \} , \\
   \gamma ~ & = (\pi \cap \rho) \setminus \gamma^* ,
\eeal
and the prefactor $\theta_{31}$ equals
\be \!\!
   \theta_{31} = \left\{
   \begin{aligned}
    & 0
      \qquad \qquad \qquad \qquad \qquad \qquad \qquad \qquad \qquad\:\,\qquad
      \text{ if } \sigma \cap \tau \neq \emptyset , \\
    & s_{3\pi4}
      \qquad \qquad \qquad \qquad \qquad \qquad \qquad \qquad \qquad\;\!\quad
      \text{ if } \{\pi\} = \{\tau\} \Leftrightarrow \{\rho\} = \{\sigma\} , \\
    & \Big( \!\!\!\!\!\!\!
            \sum_{\substack{ j \in \{3,\pi_1,\dots,\pi_{a-1}\}\\~~~\;\:
                              \cap \{\beta_{b+1},\dots,\beta_{n-3},3\},\\
                             \pi_a = \beta_b = \gamma^*_1 }}\!\!\!\!\!\!\!
            s_{\gamma^*_1 j}
      \Big)
      \Big( \!\!\!\!\!\!\!
            \sum_{ \substack{ j \in \{1,\rho_1,\dots,\rho_{b_i-1}\}\\~~~~\:
                   \cap \{\alpha_{a+1},\dots,\alpha_{n-3},1\},\\
                   \rho_b = \alpha_a = \gamma^*_{|\gamma^*|} }}\!\!\!\!\!\!\!
            s_{\gamma^*_{|\gamma^*|} j}
      \Big)
      \qquad \text{ otherwise. }
   \end{aligned} \right. \!\!\!\!\!\!\!
\label{KLTk2compactprefactor}
\ee
\ese
We have checked these three $k=2$ formulae through eight points.
\Eqn{KLTk2flip} mimicking the cases of $k=0,1$
was also proven for any multiplicity in \rcite{Brown:2018wss},
whereas the latter two reveal a much richer structure
than seen in the flavorless case.
The seemingly most complicated one, \eqn{KLTk2compact}, actually contains
much fewer nonzero terms than the other two.
Indeed, the prefactor~\eqref{KLTk2compactprefactor} vanishes
for a growing fraction of permutations, starting from $1/4$ for $n=5$
and reaching $63\%$ for $n=10$.
This means that using this double copy
to evaluate gravitational amplitudes
must be numerically more efficient at higher points.

\subsection{Six-matter double copies at low multiplicity}
\label{sec:KLTk3}

\begin{table*}[t]
\centering
\begin{tabular}{|ll|r|}
\hline
$\alpha$ & $\beta$ &
$4S[\alpha,\beta]$ \\
\hline
   $\o{3}\u{4}\;\o{5}\u{6}$ &
   $\o{3}\u{4}\;\o{5}\u{6}$ &
   $ s_{156}\,s_{256} (s_{12}\!+\!s_{34}\!+\!s_{56}) $ \\
   $\o{3}\u{4}\;\o{5}\u{6}$ &
   $\o{5}\u{6}\;\o{3}\u{4}$ &
   $-s_{156} (s_{156} (s_{12}\!+\!s_{34}\!+\!s_{56}) -4 s_{34} s_{56}) $ \\
   $\o{3}\u{4}\;\o{5}\u{6}$ &
   $\o{3}\;\o{5}\u{6}\;\u{4}$ &
   $ s_{156} (s_{256} (s_{12}\!+\!s_{34}\!+\!s_{56})
            + s_{456} (s_{12}\!+\!s_{34}\!-\!s_{56})) $ \\
   $\o{3}\u{4}\;\o{5}\u{6}$ &
   $\o{5}\;\o{3}\u{4}\;\u{6}$ &
   $ s_{156} (s_{256} (s_{12}\!+\!s_{34}\!+\!s_{56})
            + s_{126} (s_{12}\!-\!s_{34}\!+\!s_{56})) $ \\
   $\o{3}\;\o{5}\u{6}\;\u{4}$ &
   $\o{3}\;\o{5}\u{6}\;\u{4}$ &
   $ (s_{12}\!+\!s_{34}\!+\!s_{56})
     (s_{156} s_{256} - s_{356} s_{456}) $ \\
   $\o{3}\;\o{5}\u{6}\;\u{4}$ &
   $\o{5}\;\o{3}\u{4}\;\u{6}$ &
   $ - s_{256} (s_{256} (s_{12}\!+\!s_{34}\!+\!s_{56}) - 4s_{34} s_{56})
     - s_{345} s_{256} (s_{12}\!-\!s_{34}\!+\!s_{56}) $ \\
 &&$ - s_{256} s_{356} (s_{12}\!+\!s_{34}\!-\!s_{56})
     - s_{345} s_{356} (s_{12}\!-\!s_{34}\!-\!s_{56}) $ \\
\hline
\end{tabular}
\caption[a]{\small The momentum kernel for the gravitational
amplitude $M_{n=6,k=3}$. Only the six key entries are shown,
the other ten are obtained by relabeling.}
\label{tab:KLTk3}
\end{table*}

All-multiplicity gauge-invariant double-copy formulae
for more than two matter particle pairs
have proven highly non-trivial to derive.
In \tab{tab:KLTk3} we present the six-point momentum kernel
for three matter pairs.
The kinematic expressions given there should be used in the formula
\be
   M_{n,k=3} = A(\o{1},\u{2},\alpha) \tilde{A}(\u{2},\o{1},\beta) S[\alpha,\beta] \, .
\ee
We have obtained a seven-point momentum kernel for the basis combination
$A(\o{1},\u{2},\o{q},\dots) \times A(\u{2},\o{1},\o{q}',\dots)$ as well.
Some of its entries are lengthy, so we do not display them here.
However, as in the six-point case, the seven-point double-copy formula
is explicitly local and gauge-invariant,
and we provide them both in a computer-readable format in two ancillary files~\cite{Aux}.

\section{Summary and Discussion}
\label{sec:outro}

In this paper, we have elaborated on a general framework for double copies applied to gravitational theories with flavored matter and spin. The framework is based on the results and lessons learned from the extension of the color-kinematics duality to fundamental matter~\cite{Johansson:2014zca}, and in particular to QCD with massive quarks~\cite{Johansson:2015oia}. However, we expect it to be more generally applicable to gravitational amplitudes obtained from a pair of gauge theories that obey color-kinematics duality and that have massive matter transforming in non-adjoint representations of the gauge group. Due to this generality, it should also be useful for more systematic studies related to classical scattering of multiple gravitational sources and associated gravitational wave emission, as demonstrated by Shen~\cite{Shen:2018ebu}.
See also \rcites{Goldberger:2016iau,Goldberger:2017frp,Luna:2017dtq,Goldberger:2017ogt,Goldberger:2017vcg,Cheung:2018wkq,Li:2018qap,Carrillo-Gonzalez:2018pjk,Kosower:2018adc,Plefka:2018dpa,Bern:2019nnu,Plefka:2019hmz, Bautista:2019tdr,Maybee:2019jus,Guevara:2019fsj}.

Specifying the properties that identify the gravitational theory given by the double copy is in general a non-trivial problem.
The exception is when one can rely on uniqueness properties such as (super)symmetries and other easily identifiable traits of the underlying gauge theories.  A major focus of the current work is to consider the double copy obtained from QCD with $N_f$ massive quarks, which is a less straightforward task than previous identifications and classifications (see \eg \rcites{Chiodaroli:2014xia,Chiodaroli:2015wal,Chiodaroli:2014xia,Azevedo:2018dgo,Chiodaroli:2017ehv}).  
The resulting gravitational theory contains $N_f$ massive vector bosons
minimally coupled to gravity. Such non-self-interacting massive vectors are useful models for spinning black holes, 
which in the classical limit of scattering amplitudes can be used to obtain the lower-multipole corrections to gravitational observables~\cite{Kosower:2018adc,Maybee:2019jus}.
It is an automatic feature of our QCD construction that the resulting double-copy gravitational theory can at most be quadratic in the massive vectors. 

For every massive vector there is a companion scalar of the same mass, which interacts highly non-linearly with itself and with other matter field such as the (massless) axiodilaton scalar. This feature is an echo of the supersymmetric cousins of the same construction~\cite{Chiodaroli:2015wal}, which has non-linear scalar interactions parametrized by supergravity moduli spaces.
This makes it harder to use a brute force method to construct the full scalar dependence in the Lagrangian.
Nevertheless, we have through the double copy constrained the gravitational Lagrangian up to quartic couplings; the result is given in \eqn{PertLagrangian}.  In principle, the Lagrangian can be constructed to any order in $\kappa$, up to field redefinitions in the scalar fields, by matching with the higher-point amplitudes obtained from the double copy.
However, this is a tedious procedure, and it would likely be better
to pinpoint the remaining scalar interactions
using some geometrical consideration that would yield
a closed form for the metric that parametrizes the scalar space.

From the perspective of using the current construction for classical calculations of relevant observables, the massive scalars would most likely obfuscate any physical process. One way to avoid dealing with such scalars is to consider a double copy where the spinning matter belongs to only one gauge theory, and the other gauge theory has only scalar matter, as done in \rcite{Maybee:2019jus}. The general framework of this paper is well suited for such a construction, and in particular problems with multiple sources may strongly benefit from  the considerations in this work. Having the massive spin degrees of freedom come exclusively from one of the two gauge-theories should also make the problem of identifying the gravitational theory more straightforward. 

A second major part of this paper was to elaborate on general features of the double copies with flavored matter --- agnostic of the type of theories being considered --- and to give manifestly gauge-invariant expressions
for the double-copy formulae.
Doing so, we found new all-multiplicity formulae for
a gauge-invariant double copy for amplitudes with two flavored pairs
of matter particles, as well as partial results for three matter pairs.
In particular, we provide a double-copy formula~\cite{Aux}
for a seven-point amplitude with three matter pairs.
In the context of classical gravitational radiation, such a seven-point amplitude contains the information of the same post-Minkowskian order as considered in~\rcite{Shen:2018ebu}. A remaining task is to find the general formulae for the gauge-invariant double copy, with three or more flavors. While these general formulae would be useful for classical calculations, they might contain more information than what is necessary for practical calculations, as the classical limit usually involves placing various particles on shell, where the double copy and the corresponding momentum kernel factorizes into smaller pieces. Understanding this more systematically should be a promising research direction.

\paragraph{Double-copy tension at higher spins.}

Coming back to the double copy of massive spinning particles,
we recall the higher-spin issue~\cite{Arkani-Hamed:2017jhn}
that we encountered in \sec{sec:4pt}.
Namely, in the general-spin formulae~\eqref{comptonQCD} and \eqref{comptonGR1}
the power of the factor~$\bra{3}1|2]$ becomes negative for
$s>1$ and $s>2$, respectively. In that case the amplitudes contain unphysical poles. If so, the BCFW-shift variable must not fall off fast enough at infinity for $s\ge3/2$ in gauge theory. The $\bra{3}1|2]$ pole can be removed if the neglected boundary contribution is recovered by other means. \Rcites{Arkani-Hamed:2017jhn,Chung:2018kqs}
treated this as an invitation to find an additional amplitude contribution
that vanishes on all of the physical poles
and subtracts out the unphysical pole from the BCFW result~\eqref{comptonQCD}.
Such a contribution, however, spoils the high-energy/massless limit as the mass parameter $m$ must appear explicitly in the denominator to compensate for the engineering dimension of the interaction.
Indeed, as argued in \rcite{Arkani-Hamed:2017jhn},
the singular high-energy limit of such amplitudes is consistent with the expected composite nature of higher-spin particles.

Let us now add a new perspective to this puzzle. At higher spins there seems to also be a tension between the double copy and locality. Indeed, consider the double copy~\eqref{comptonDC4pt} of the gauge-theory amplitudes
with spins $s_1=1$ and $s_2=1/2$.
They both contain only physical poles and satisfy color-kinematics duality,
so we have a consistent double copy that gives a local gravitational amplitude
\beal
\label{comptonGR1o12}
  M(1^{\{a\}}\!,2^+\!,3^-\!,4^{\{b\}}) &
    = \frac{i\bra{3}1|2]}{(s_{12}\!-\!m^2)(s_{13}-m^2)s_{23}} 
      \big( [1^a 2] \braket{3\;\!4^b} + \braket{1^a 3} [2\;\!4^b]
      \big)^{\odot 3}\,,
\eeal
for a massive Rarita-Schwinger field (which has the interpretation of a massive gravitino in a gravitational theory with broken supersymmetry~\cite{Chiodaroli:2017ehv,Chiodaroli:2018dbu}).
At the same time, let us apply the double copy~\eqref{comptonDC4pt}
to gauge theories with spins $s_1=3/2$ and $s_2=0$, respectively. In order to reproduce the above amplitude via
\fontsize{10.5pt}{0}\selectfont
\be
\label{comptonGR32o0}
  M(1^{\{a\}}\!,2^+\!,3^-\!,4^{\{b\}})\;\!\!
    =\;\!\!-is_{23} A(1^{\{a\}}\!,2^+\!,3^-\!,4^{\{b\}}) A(1,3^-\!,2^+\!,4)\;\!\!
    =\;\!\!\frac{\bra{3}1|2]^{2}}{(s_{13}\!-\!m^2)}
      A(1^{\{a\}}\!,2^+\!,3^-\!,4^{\{b\}}) ,
\ee
\normalsize
the gauge-theory amplitude with the spin-3/2 particle must contain a spurious pole:
\be
   A(1^{\{a\}}\!,2^+\!,3^-\!,4^{\{b\}})
    = \frac{ i \big( [1^a 2] \braket{3\;\!4^b} + \braket{1^a 3} [2\;\!4^b]
               \big)^{\odot 3} }
           { (s_{12}\!-\!m^2) s_{23} \bra{3}1|2] } \, .
\ee
If we follow \rcites{Arkani-Hamed:2017jhn,Chung:2018kqs}
and attempt to cancel out the pole by additional contributions free of physical poles,
this will likely non-trivially interfere with color-kinematics duality
and therefore the double copy.
Even if one manages to find a color-dual gauge-theory amplitude for massive spin-3/2 particles, it must introduce 
a correction to the double copy (\ref{comptonGR32o0}) such that it becomes seemingly impossible to manifest the trivial identity $s=1+1=3/2+1/2=2+0$ by various partitioning of the spins on the two sides of the double copy. 

Having pointed out the above potential problems for higher-spin double copies, let us end on a positive note. We know that the double copy is an intrinsic feature of string theory. Not only does it follow from the work of KLT~\cite{Kawai:1985xq}, but also more recent results show that the double-copy structure 
permeates all types of strings: the bosonic, heterotic, closed and open superstrings \cite{Mafra:2011nw,Schlotterer:2012ny,Broedel:2013tta,Ochirov:2013xba,Stieberger:2014hba,Huang:2016tag,Azevedo:2018dgo}. As string theory contains a multitude of higher-spin massive particles we can conclude that some version of a double copy must exist for these particles. While the infinite tower of massive modes may non-trivially conspire and cause technical challenges, there are good reasons to expect that one can isolate individual modes by investigating the residues of massive poles in the tree-level amplitudes. Indeed, it would be interesting to initiate a systematic study using string theory and see if the modes can be isolated without disturbing the double-copy structure. We leave this question and those raised above for the future.

\acknowledgments

We would like to thank John Joseph Carrasco, Marco Chiodaroli, Rea Dalipi,
Guillermo Garc\'ia Fern\'andez, Alfredo Guevara, Murat G\"{u}naydin, David Kosower, Gregor K\"{a}lin, Ben Maybee, Gustav Mogull, Donal O'Connell, Radu Roiban, Chia-Hsien Shen,  Nils Siemonsen, David Skinner, Fei Teng and Justin Vines for stimulating discussions
and collaboration on related topics.
The research of H.J. is supported by the Swedish Research Council under grant 621-2014-5722, the Knut and Alice Wallenberg Foundation under grant KAW 2013.0235, and the Ragnar S\"{o}derberg Foundation (Swedish Foundations' Starting Grant).
A.O.'s research is funded by the European Union's Horizon 2020 research and innovation programme under the Marie Sk{\l}odowska-Curie grant agreement 746138.

\appendix
\section{Axiodilaton gravity in four dimensions}
\label{app:fatgravity}

Here we review the massless gravitational theory obtained as a double copy
of pure Yang-Mills theory.
The string-theoretic origin of the KLT double copy~\cite{Kawai:1985xq}
implies that this theory
(sometimes referred to as ${\cal N}=0$ or NS-NS gravity~\cite{Geyer:2017ela})
is Einstein's gravity coupled to a dilaton scalar $\phi$
and a Kalb-Ramond two-form $B_{\mu\nu}$.
Its Lagrangian is given \eg in \rcite{Polchinski:1998rq}:
\begin{equation}
   {\cal L}_0 = \frac{2}{\kappa^2} \bigg( {-R}
    + \frac{4}{D-2} \partial_\mu \phi\;\!\partial^\mu \phi
    + \frac{1}{12} e^{-8\phi/{(D-2)}} H_{\lambda\mu\nu} H^{\lambda\mu\nu}
      \bigg) ,
\label{MasslessGravityLagrangian}
\end{equation}
where
$H_{\la\mu\nu}
=\partial_\la B_{\mu\nu} + \partial_\mu B_{\nu\la} + \partial_\nu B_{\la\mu}$
is the fully antisymmetric field strength for $B_{\mu\nu}$.
The double-copy relation of this theory to pure Yang-Mills in a classical setting
was investigated in \rcites{Goldberger:2016iau,Luna:2016hge}.

In four dimensions, it is convenient to dualize
the two-form $B_{\mu\nu}$ to an axion pseudoscalar $a$
through a type of the Legendre transform
\be
   {\cal L}_\text{axion} = \frac{1}{6\kappa^2}
      \Big( e^{-4\phi}  H_{\la\mu\nu} H^{\la\mu\nu}\!
          - 4 H_{\la\mu\nu} E^{\la\mu\nu\rho} \partial_\rho a
      \Big) ,
\label{axionlagrangianlegendre}
\ee
where the Levi-Civita tensors are defined through the Levi-Civita symbol as
\be
   E_{\la\mu\nu\rho} = \sqrt{-g}\;\!\epsilon_{\la\mu\nu\rho} , \qquad
   E^{\la\mu\nu\rho}
    = g^{\la\alpha} g^{\mu\beta} g^{\nu\gamma} g^{\rho\delta}
      E_{\alpha\beta\gamma\delta}
    = \frac{\epsilon^{\la\mu\nu\rho}}{\sqrt{-g}} , \qquad
   \epsilon_{0123}=-\epsilon^{0123}=1 .
\label{LeviCivita}
\ee
Now the field equation for $a$
sets $\epsilon^{\la\mu\nu\rho} \partial_\la H_{\mu\nu\rho}$ to zero,
which means that $H$ is a closed and therefore exact three-form.
Having thus guaranteed $H = \d B$ on the equations of motion,
we can now consider $H_{\la\mu\nu}$ as a separate field.
Its field equations are
\be
   H_{\la\mu\nu}\! = 2e^{4\phi} E_{\la\mu\nu\rho} \partial^\rho a
   \qquad \Leftrightarrow \qquad
   H^{\la\mu\nu}\! = 2e^{4\phi} E^{\la\mu\nu\rho} \partial_\rho a .
\label{axiondualization}
\ee
Then for both terms in the axion Lagrangian~\eqref{axionlagrangianlegendre} we have
\begin{subequations}
\begin{align}
   e^{-4\phi} H_{\la\mu\nu} H^{\la\mu\nu} &
      = 4e^{4\phi} \epsilon_{\la\mu\nu\rho} \epsilon^{\la\mu\nu\sigma}
        \partial^\rho a\;\!\partial_\sigma a
      = -24e^{4\phi} \partial_\mu a\;\!\partial^\mu a , \\*
  -4H_{\la\mu\nu} E^{\la\mu\nu\rho} \partial_\rho a &
      =-8e^{4\phi} \epsilon_{\la\mu\nu\rho} \epsilon^{\la\mu\nu\sigma}
        \partial^\rho a\;\!\partial_\sigma a
      = 48e^{4\phi} \partial_\mu a\;\!\partial^\mu a ,
\end{align}%
\end{subequations}
so that the full Lagrangian~\eqref{MasslessGravityLagrangian}
becomes equivalent to
\be
   {\cal L}_0 = \frac{2}{\kappa^2} \Big( {-R}
    + 2 g^{\mu\nu} \big( \partial_\mu \phi\;\!\partial_\nu \phi
    + e^{4\phi} \partial_\mu a\;\!\partial_\nu a \big)
      \Big) .
\label{MasslessGravityLagrangian4d}
\ee

At this point, we could rescale the dilaton and axion by $\kappa/(2\sqrt{2})$
to fix the normalization of their kinetic terms to $1/2$
and deal with a perturbation theory of two real scalars.
The four-dimensional double copy, however, makes it preferable
to introduce a complex scalar field~\cite{deRoo:1984zyh,Bergshoeff:1985ms}
\be
   Z = \frac{2a + i(e^{-2\phi}-1)}{2a + i(e^{-2\phi}+1)} ,
\label{complexdilaton}
\ee
which we refer to as the axiodilaton.
This change of variables produces the following Lagrangian
(see \eg \cite{Schwarz:1992tn})
\beal
   {\cal L}_0 & = \frac{2}{\kappa^2}
      \bigg({-R} + 2 g^{\mu\nu} \frac{ \partial_\mu \bar{Z}\;\!\partial_\nu Z }
                                     { (1-\bar{Z}Z)^2 }
      \bigg) .
\label{MasslessGravityLagrangian4d2}
\eeal
Its perturbative expansion in terms of Feynman rules
is obtained by rescaling $Z \rightarrow \kappa Z/2$
and expanding the series into
a tower of exclusively two-derivative matter interactions:
\beal
   \sqrt{-g} {\cal L}_0 &
    = -\frac{2}{\kappa^2} \sqrt{-g} R
    + (\eta^{\mu\nu} - \kappa \hh^{\mu\nu}) \partial_\mu \bar{Z}\;\!\partial_\nu Z
      \bigg( 1 + \sum_{k=1}^\infty
               \frac{\kappa^{2k}}{2^{2k}} (k+1) (\bar{Z}Z)^k \bigg) \\ &
    = -\frac{2}{\kappa^2} \sqrt{-g} R
    + \partial_\mu \bar{Z}\;\!\partial^\mu Z
    - \kappa h^{\mu\nu} \partial_\mu \bar{Z}\;\!\partial_\nu Z
    + \frac{\kappa^2\!}{2} \bar{Z}Z
      \partial_\mu \bar{Z}\;\!\partial^\mu Z + O(\kappa^3) .
\label{MasslessGravityLagrangianPT}
\eeal

\section{Massive Majorana fermions}
\label{app:majorana}

In most of this paper we deal with Dirac and Majorana fermions simultaneously.
However, since most physicists are more accustomed to the former
than the latter~\cite{Pal:2010ih},
this appendix serves to remove any potential confusion
in the simple transition between the two.

As the starting point we take the textbook formulae~\cite{Peskin:1995ev}
for Dirac quantum fields,
which we write in the little-group-covariant way as
\begin{subequations} \begin{align}
   \Psi(x) & =\!\sum_{a=1,2} \int\!\!\frac{\dd^3p}{2p^0}
      \Big[ e^{-i p \cdot x} u^a_p c_a(\vec{\:\!p})
          + e^{i p \cdot x} v_{p\:\!a} d^{\dagger a}(\vec{\:\!p})
      \Big] \Big|_{p^0=\sqrt{\vec{\:\!p}^2+m^2}} , \\
   \bar{\Psi}(x) & =\!\sum_{a=1,2} \int\!\!\frac{\dd^3p}{2p^0}
      \Big[ e^{-i p \cdot x} \bar{v}^a_p d_a(\vec{\:\!p})
          + e^{i p \cdot x} \bar{u}_{p\:\!a} c^{\dagger a}(\vec{\:\!p})
      \Big] \Big|_{p^0=\sqrt{\vec{\:\!p}^2+m^2}}
    = \Psi^\dagger(x) \gamma^0 ,\!\!\!\!\!\!\!
\end{align} \label{freefields}%
\end{subequations}
where the fermion annihilation and creation operators obey
the non-vanishing anticommutation rules
$ \{ c_a(\vec{\:\!p}) , c^{\dagger b}(\vec{\:\!q}) \}
= 2p^0 \delta_a^b \del^{\,(3)}(\vec{\:\!p}-\:\!\!\vec{\:\!q}) $ and
$ \{ d_a(\vec{\:\!p}) , d^{\dagger b}(\vec{\:\!q}) \}
= 2p^0 \delta^a_b \del^{\,(3)}(\vec{\:\!p}-\:\!\!\vec{\:\!q}) $.
One can then build up the perturbation theory
in the usual way using these fields as building blocks.
For example, they satisfy the Dirac equations
$(i\!\!\not{\!\partial}-m)\Psi
=\bar{\Psi}(i\!\overleftarrow{\!\!\!\not{\!\partial}\!\!\!}+m)=0$
and the equal-time anticommutation relation
$ \{ \Psi(t,\vec{\:\!x}) , \bar{\Psi}(t,\vec{\:\!y}) \}
= \gamma^0 \delta^{(3)}(\vec{\:\!x}-\!\vec{\:\!y}) $.

Now the transition to Majorana spinors amounts to
identifying the operators $d_a(\vec{\:\!p})$ with $c_a(\vec{\:\!p})$.
This reduces the number of degrees of freedom of the resulting quantum field
$\Psi_{\rm M}(x)$ by two due to the fact that it obeys
an additional reality condition $\Psi_{\rm M}=\Psi_{\rm M}^{\rm c}$
in \eqn{MajoranaRealityCondition}.
Indeed, it is straightforward to check it using
the conjugation properties~\eqref{diracconjugation}
for the external spinors $u_p^a$ and $v_p^a$.
Their interpretation as wavefunctions for the one-particle states
appearing in the Feynman rules changes only slightly
between the two cases:
\begin{subequations} \begin{align}
   \begin{aligned}
   \braket{0|\Psi(x)|\psi(\vec{\:\!p},a)} &
    = \braket{0|\Psi_{\rm M}(x)|\psi_{\rm M}(\vec{\:\!p},a)}
    = u^a_{p}\,e^{-i p \cdot x} , \qquad \quad
   \braket{0|\Psi(x)|\bar{\psi}(\vec{\:\!p},a)} = 0 , \\*
   \braket{0|\bar{\Psi}(x)|\bar{\psi}(\vec{\:\!p},a)} &
    = \braket{0|\bar{\Psi}_{\rm M}(x)|\psi_{\rm M}(\vec{\:\!p},a)}
    = \bar{v}^a_{p}\,e^{-i p \cdot x} , \qquad\:\!\quad
   \braket{0|\bar{\Psi}(x)|\psi(\vec{\:\!p},a)} = 0 ,
   \end{aligned} \\*
   \begin{aligned}
   \braket{\psi(\vec{\:\!p},a)|\bar{\Psi}(x)|0} &
    = \braket{\psi_{\rm M}(\vec{\:\!p},a)|\bar{\Psi}_{\rm M}(x)|0}
    = \bar{u}_{p\:\!a}\,e^{i p \cdot x} , \qquad \quad
   \braket{\bar{\psi}(\vec{\:\!p},a)|\bar{\Psi}(x)|0} = 0 , \\*
   \braket{\bar{\psi}(\vec{\:\!p},a)|\Psi(x)|0} &
    = \braket{\psi_{\rm M}(\vec{\:\!p},a)|\Psi_{\rm M}(x)|0}
    = v_{p\:\!a}\,e^{i p \cdot x} , \qquad\:\!\quad
   \braket{\psi(\vec{\:\!p},a)|\Psi(x)|0} = 0 ,
   \end{aligned}
\end{align} \label{wavefunctions}%
\end{subequations}
where all energies $p^0=\sqrt{\vec{\:\!p}^2+m^2}$ are taken positive.
For instance, in the complex case
the expectation value $\braket{0|\bar{\Psi}(x)|\psi(\vec{\:\!p},a)}$
with an electron/quark state vanishes,
but in the real case of a neutrino state it gives $\bar{v}^a_{p}\,e^{-i p \cdot x}$,
as shown above.

Under the Majorana projection, the Dirac Lagrangian~\eqref{DiracLagrangian} becomes
\be
\label{MajoranaLagrangian}
   {\cal L}_\text{Majorana}
    = \frac{1}{2} \bar{\Psi}_{\rm M} (i\gamma^\mu D_\mu - m) \Psi_{\rm M} .
\ee
The resulting three-point vertex is composed of two possible contractions
\begin{align}
\label{GaugeVertexMajorana}
   {\cal L}_{\bar{\Psi} \Psi A} &
    = \frac{g}{2} \bar{\Psi}_{\rm M}\!\not{\!\!A} \Psi_{\rm M} \\* &
   \Rightarrow
   \scalegraph{0.9}{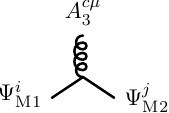}\!\!\!= \frac{1}{2}\!
   \left[\scalegraph{0.9}{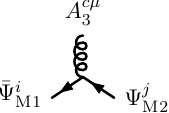}\!\!-\!\!\scalegraph{0.9}{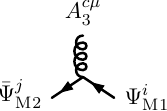}\right]\!
    = \frac{i g}{2\sqrt{2}} \tf^{icj}
      [\gamma^\mu \oplus \gamma^{\mathsf{T}\mu}]. \nn
\end{align}
where for simplicity we assumed the adjoint representation of the fermion
and used the antisymmetry of the structure constants.

\section{Gravitational coupling of Dirac fermion}
\label{app:gravitydirac}

In \sec{sec:v3spin1half} we used
the three-point vertex~\eqref{GravityVertexDirac}
with which a Dirac particle interacts gravitationally at lowest order in $\kappa$.
Let us here derive its on-shell part,
assuming that the end result must look like
\be
   (\sqrt{-g}{\cal L})_{\bar{\Psi} \Psi \hh}
    =-\frac{i\kappa}{2} \hh_{\mu\nu} T^{\mu\nu}
    + \frac{i\kappa}{4} \hh\;\!T^\la_{~\,\la}
    = -\frac{i\kappa}{2} h_{\mu\nu} T^{\mu\nu} + {\cal O}(\kappa^2) .
\label{GravityVertexDirac2}
\ee
Here the linearized stress-energy tensor $T^{\mu\nu}$
should be symmetric and conserved on the equations of motion,
$\partial_\mu T^{\mu\nu} \simeq \partial_\nu T^{\mu\nu} \simeq 0$.
We wish to obtain it from the Noether energy-momentum tensor in flat space
\be
   T^{\mu\nu}_\text{N}
    = \frac{\partial{\cal L}_\text{Dirac}}{\partial(\partial_\mu \Psi)}
      \partial^\nu \Psi
    + \partial^\nu \bar{\Psi}
      \frac{\partial{\cal L}_\text{Dirac}}{\partial(\partial_\mu \bar{\Psi})}
    - \eta^{\mu\nu} {\cal L}_\text{Dirac}
    \simeq i \bar{\Psi} \gamma^\mu \partial^\nu \Psi , \qquad \quad
   \partial_\mu T^{\mu\nu}_\text{N} \simeq 0 ,
\ee
which is not symmetric.
A known remedy for that is to add a Belinfante contribution~\cite{1940Phy.....7..449B,Rosenfeld:1940}
\be
   T^{\mu\nu} = T^{\mu\nu}_\text{N}
    + \partial_\lambda B^{\lambda\mu\,\nu} , \qquad \quad
   B^{\lambda\mu\,\nu} = \frac{1}{2}
      \big[ S^{\lambda\,\mu\nu}_\text{N} + S^{\mu\,\nu\lambda}_\text{N}
          - S^{\nu\,\lambda\mu}_\text{N} \big] ,
\ee
that is constructed using the Noether intrinsic spin tensor
\be
   S^{\lambda\,\mu\nu}_\text{N} = -i
      \bigg[ \frac{\partial{\cal L}_\text{Dirac}}{\partial(\partial_\la \Psi)}
             S^{\mu\nu}_{s=1/2} \Psi
           - \bar{\Psi} S^{\mu\nu}_{s=1/2}
             \frac{\partial{\cal L}_\text{Dirac}}{\partial(\partial_\la \bar{\Psi})}
      \bigg]
    = \frac{i}{2} \bar{\Psi} \gamma^\la \gamma^{[\mu} \gamma^{\nu]} \Psi .
\ee
The needed properties of the resulting tensor $T^{\mu\nu}$ follow from
the conservation of the total angular momentum tensor
$ J^{\lambda\,\mu\nu}_\text{N}
= x^\mu T^{\lambda\nu}_\text{N} - x^\nu T^{\lambda\mu}_\text{N}
+ S^{\lambda\,\mu\nu}_\text{N} $.
Explicitly, the stress-energy tensor becomes
\be
   T^{\mu\nu} \simeq \frac{i}{2}
      \big[ \bar{\Psi} \gamma^{(\mu} \partial^{\nu)} \Psi
          - (\partial^{(\mu} \bar{\Psi}) \gamma^{\nu)} \Psi \big] .
\ee
Coupling it to the linearized graviton via \eqn{GravityVertexDirac2},
we arrive at the interaction vertex~\eqref{GravityVertexDirac}.

\section{Cheung-Remmen gravitational vertices}
\label{app:gravityvertices}

In this appendix we summarize the perturbation theory
for general relativity formulated by Cheung and Remmen in \rcite{Cheung:2017kzx}.
They discovered that after a suitable gauge fixing the Einstein-Hilbert action
can be rewritten in a form, in which the infinite tower of gravitational vertices
is generated by a single geometric series:
\begin{align}
\label{CheungRemmenAction}
   S_\text{EH} & + S_\text{GF}
    = -\frac{2}{\kappa^2}\!\int\!d^4 x \sqrt{-g} R + S_\text{GF} \\ &
    = \int\!d^4 x \bigg(
      2 \partial_{[\rho} \hh^{\mu\rho} \partial_{\sigma]} \hh^{\nu\sigma}\!
    + \frac{1}{2} \hh^{\rho\sigma} \partial_\rho \partial_\sigma \hh^{\mu\nu}\!
    - \frac{1}{2} \partial_\rho \hh^{\rho\sigma} \partial_\sigma \hh^{\mu\nu}\!
    - \frac{1}{2} \eta_{\rho\sigma} \hh^{\mu\rho} \partial^2 \hh^{\nu\sigma} 
      \bigg)\!\sum_{k=0}^\infty \kappa^k \hh^k_{\mu\nu} . \nn
\end{align}
Similarly to the Landau-Lifshitz formulation of general relativity~\cite{Landau:1982dva,Poisson:2014},
the main variable here is the perturbation of the ``gothic inverse metric''
\beal
   \gg^{\mu\nu} & = \sqrt{-g} g^{\mu\nu}
    = \eta^{\mu\nu} - \kappa \hh^{\mu\nu} , \\
   \gg_{\mu\nu} & \overset{\substack{\Downarrow \\ ~}}{=}
      \frac{g_{\mu\nu}}{\sqrt{-g}}
    = \sum_{k=0}^\infty \kappa^k \mathfrak{h}^k_{\mu\nu}
    = \eta_{\mu\nu} + \kappa \hh_{\mu\nu}
    + \kappa^2 \hh_{\mu\rho} \hh^{\rho}_{~\,\nu}
    + \kappa^3 \hh_{\mu\rho} \hh^{\rho}_{~\,\sigma} \hh^{\sigma}_{~\,\nu}
    + {\cal O}(\kappa^4) ,
\eeal
where the Lorentz indices are raised and lowered
with the flat background metric~$\eta_{\mu\nu}$.

At the linearized level, $\hh_{\mu\nu}$ corresponds
to the trace-reversed version $h_{\mu\nu} - h \eta_{\mu\nu} /2$
of the more conventional graviton $\kappa h_{\mu\nu} = g_{\mu\nu} - \eta_{\mu\nu}$.
The graviton propagator that follows from the action~\eqref{CheungRemmenAction} is
\be
   \scalegraph{1.0}{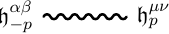} =
   \frac{i}{2p^2}
   \bigg( \eta^{\alpha\mu} \eta^{\beta\nu} + \eta^{\alpha\nu} \eta^{\beta\mu}
        - \frac{1}{2} \eta^{\alpha\beta} \eta^{\mu\nu}
        - \frac{2}{p^4} p^\alpha p^\beta p^\mu p^\nu \bigg) ,
\label{propG}
\ee
while the $k$-valent vertices are 
all given by a single explicit formula
\begin{align}
   \scalegraph{0.9}{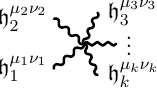}\!& = i \kappa^{k-2}
      \sum_{\sigma \in {\rm S}_k}
      K^{\mu_{\sigma_1}\:\!\!\nu_{\sigma_1}\;\mu_{\sigma_2}\nu_{\sigma_2}\;
         \mu_{\sigma_3}\nu_{\sigma_k}}(p_1,p_2)
      \eta^{\nu_{\sigma_3}\mu_{\sigma_4}} \eta^{\nu_{\sigma_4}\mu_{\sigma_5}} \dots
      \eta^{\nu_{\sigma_{k-1}}\mu_{\sigma_k}} , \nn \\
   K^{\alpha\beta\,\gamma\delta\,\mu\nu}(p,q) & = -\bigg\{\!
      \bigg( p^\alpha q^\gamma - q^\alpha p^\gamma
           - \frac{1}{2} q^2 \eta^{\alpha\gamma} \bigg)
      \eta^{\beta\mu} \eta^{\delta\nu}
    - \frac{1}{2} (p-q)^\alpha q^\beta \eta^{\gamma\mu} \eta^{\delta\nu} \bigg\} .
\label{vGgen}
\end{align}
where every pair of indices $(\mu_i,\nu_i)$ should in principle be symmetrized,
but this symmetrization can be left implicit
as long as the symmetric propagators~\eqref{propG} and
wavefunctions~$\varepsilon_{\mu\nu} = \varepsilon_{\mu}\varepsilon_{\nu}$ are used.
As a consistency check, we verified that the above vertices reproduce
the three-, four- and five-graviton amplitudes
obtained by the standard KLT relations (mimicked by \eqn{KLTk2flip}).

Moreover, this choice of variables is convenient for considering
gravitationally covariantized matter Lagrangians.
Namely, the scalar kinetic term only contrubutes
to the three-point vertex~\eqref{GravityVertexScalar},
so vertices with two scalar fields and more than one graviton
come exclusively from the mass term, which involves
\be
\label{MetricDetSqrt}
   \sqrt{-g}
     = \exp\!\bigg( {-}\frac{1}{2} \sum_{k=1}^\infty \frac{\kappa^k}{k}
                  \hh^k_{\mu\nu} \eta^{\mu\nu} \bigg)
     = 1 - \frac{\kappa}{2} \mathfrak{h}
         + \frac{\kappa^2}{8} \big( \hh^2 - 2 \hh^{\mu\nu} \hh_{\mu\nu} \big)
         + O(\kappa^3) ,
\ee
where $\hh = \hh^{\mu\nu} \eta_{\mu\nu}$.
For instance, the resulting four-point vertex is explicitly
\begin{align}
\label{GravityVertexScalar2}
   \scalegraph{0.9}{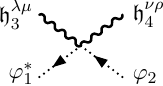}\!= \frac{i\kappa^2}{4} m^2
      \big[ \eta^{\la\nu} \eta^{\mu\rho} + \eta^{\la\rho} \eta^{\mu\nu}
          - \eta^{\la\mu} \eta^{\nu\rho} \big] . 
\end{align}

For vector bosons it is the mass term that is only contributes
to the three-point vertex~\eqref{GravityVertexVector},
and the kinetic term generates higher-point vertices,
such as the quartic vertex
\begin{align}
   \begin{aligned}
       \scalegraph{0.9}{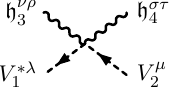} \\~ \\~
   \end{aligned} &
   \begin{aligned} = &\,i \kappa^2\!
      \bigg\{ \eta^{\la(\nu} \eta^{\rho)\mu} p_1^{(\sigma} p_2^{\tau)}
            - \eta^{\la(\nu} p_2^{\rho)} p_1^{(\sigma} \eta^{\tau)\mu} \\ & \qquad
            - \frac{1}{2} \eta^{\sigma\tau}
              \big[ (p_1 \cdot p_2) \eta^{\la(\nu} \eta^{\rho)\mu}
                  + \eta^{\la\mu} p_1^{(\nu} p_2^{\rho)}
                  - p_1^\mu \eta^{\la(\nu} p_2^{\rho)}
                  - p_2^\la \eta^{\mu(\nu} p_1^{\rho)} \big] \\ & \qquad
            + \frac{1}{8}
              \big( \eta^{\la\mu} (p_1 \cdot p_2) - p_2^\la p_1^\mu \big)
              \big( \eta^{\nu\rho} \eta^{\sigma\tau}\!
                  + \eta^{\nu\sigma} \eta^{\rho\tau}\!
                  + \eta^{\nu\tau} \eta^{\rho\sigma} \big)
      \bigg\}
   \end{aligned} \nn \\ & \quad\,
    + \big\{ (\nu\rho) \leftrightarrow (\sigma\tau) \big\} .
\label{GravityVertexVector2}
\end{align}

\bibliographystyle{JHEP}
\bibliography{references}

\end{document}